%% file: main.tex
\documentclass[10pt,twocolumn,letterpaper]{article}

\usepackage[]{cvpr}      % To produce the REVIEW version
\usepackage{tabularx}
\usepackage{booktabs}
\usepackage[table]{xcolor}
\usepackage{graphicx}
\usepackage{makecell}
\usepackage[acronym]{glossaries}
\usepackage{multirow}

\input{preamble}
\input{acronym}
\definecolor{cvprblue}{rgb}{0.21,0.49,0.74}
\usepackage[pagebackref,breaklinks,colorlinks,allcolors=cvprblue]{hyperref}

\def\confName{}
\def\confYear{}

\title{StreamAZ: Line-by-Line SAR Focusing with State Space Models}
\author{%
  Sebastian Fieldhouse\thanks{These authors contributed equally to this work.}\\
  College of Semiconductor Research\\
  National Tsing Hua University\\
  {\tt\small sebastianfieldhouse.2@gmail.com}
  \and
  Roberto Del Prete\footnotemark[1]\\
  $\Phi$-lab, European Space Agency (ESA)\\
  {\tt\small roberto.delprete@esa.int}
  \and
  Gabriele Daga\\
  $\Phi$-lab, European Space Agency (ESA)\\
  {\tt\small gabriele.daga@ext.esa.int}
  \and
  Nathaniel Rensly\\
  Department of Electrical Engineering\\
  National Tsing Hua University\\
  {\tt\small rensly2208@gmail.com}
  \and
  Gabriele Meoni\\
  Advanced Concepts and Studies Office\\
  European Space Agency (ESA)\\
  {\tt\small gabriele.meoni@esa.int}
  \and
  Kea-Tiong Tang\\
  Department of Electrical Engineering\\
  National Tsing Hua University\\
  {\tt\small kttang@ee.nthu.edu.tw}
}

\begin{document}

\twocolumn[{%
\renewcommand\twocolumn[1][]{#1}%
\maketitle
\begin{center}
    \centering
    \includegraphics[width=0.95\linewidth,
    trim=0.1cm 2cm 0.1cm 0.1cm,
    clip]{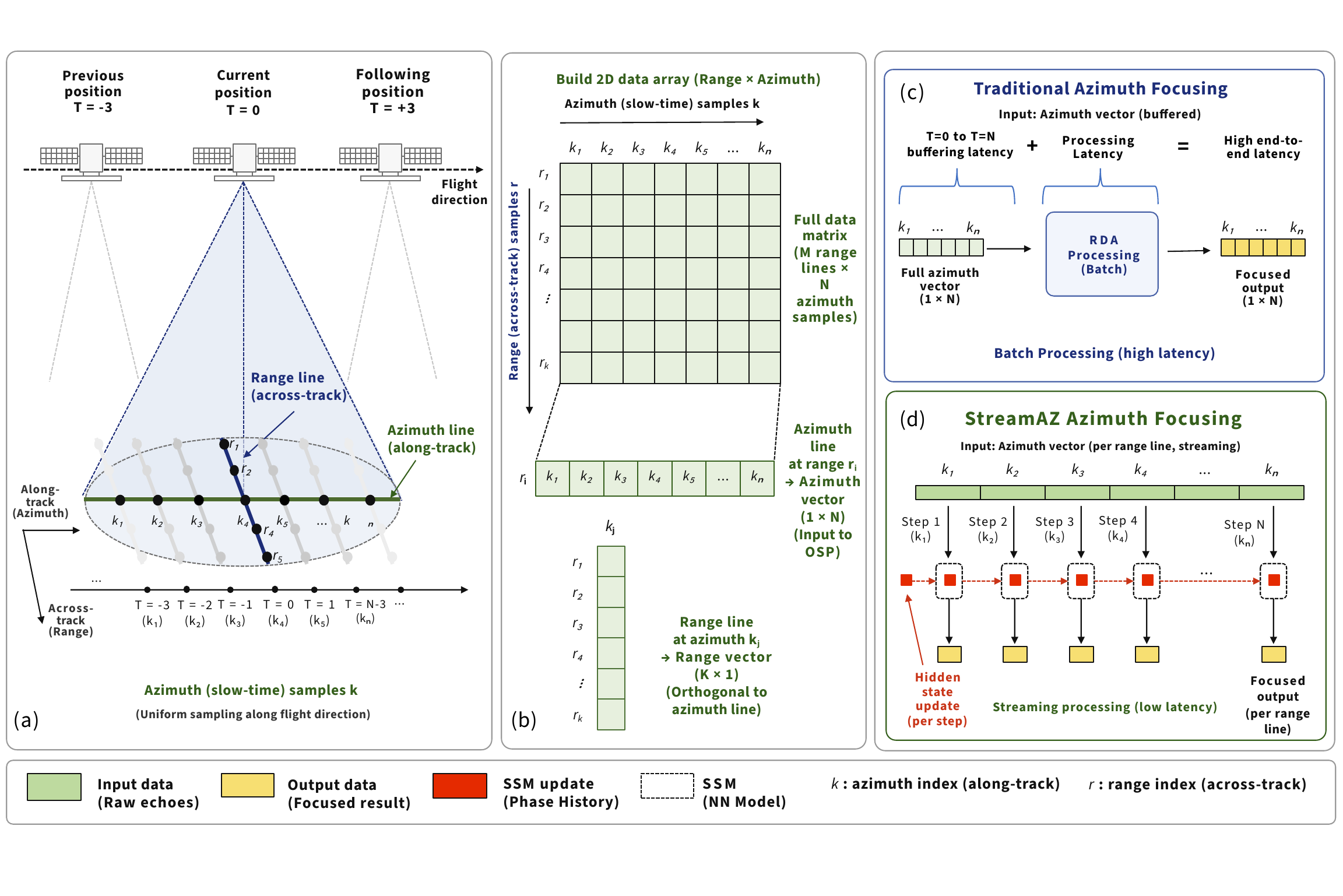}
    \captionof{figure}{SAR image formation with the StreamAZ framework.
  (a) Platform samples the scene sequentially along slow time, indexed by
  pulse/azimuth index \(k\), while each received pulse contains fast-time samples
  over range bins \(r\).
  (b) Stacking pulses yields a complex phase-history matrix whose rows are input
  range vectors \(z_{k,:}\) and whose columns are azimuth sequences
  \(z_{1:k,r}\) at fixed range bin.
  (c) Conventional RDA-style azimuth focusing requires the aperture support needed
  by the matched-filter operations before corresponding focused output can
  be formed; its end-to-end latency therefore includes aperture-buffering delay
  plus compute time.
  (d) StreamAZ replaces learned azimuth-focusing stage with a causal state-space
  recurrence applied independently to each range-bin sequence.  At pulse \(k\),
  the current sample \(z_{k,r}\) and range feature \(p_r\) update a
  range-specific hidden state,
  \(s_{k,r}=g_{\theta}(s_{k-1,r},z_{k,r},p_r)\), followed by the shared readout
  \(\hat{x}_{k,r}=h_{\theta}(s_{k,r})\).  The parameters are shared across all
  range bins, while the hidden states remain range-bin specific.  Running this
  recurrence for all \(r\) emits the focused row \(\hat{x}_{k,:}\) as new pulses
  arrive.%, using fixed per-range-bin state rather than a full azimuth-aperture
  %buffer in the learned focusing stage.
  }\label{fig:intro-online-processing}
\end{center}%
}]

\begin{abstract}
Conventional focusing methods for Synthetic Aperture Radar (SAR) employ block-wise processing efficiently but remain latency-heavy processes that prevent the realisation of a closed-loop cognitive SAR system. We present an online SAR image-formation framework,  StreamAZ, that treats sensing as a stream and produces focused images in a line by line fashion. Our method uses a tiny state-space surrogate model trained with teacher-student distillation and multi-stage losses. We evaluate the method on real SAR data from Maya4, a Sentinel-1-derived dataset containing raw, range-compressed, range-cell-migration-corrected, and azimuth-compressed products. Relative to a linewise digital-signal-processing baseline, our framework delivers approximately 70$\times$ lower latency and 130$\times$ lower memory use; on a single AMD CPU core it processes one row in 16 ms with a memory footprint of 6 MB whilst maintaining a focusing quality high enough to support downstream decisions, which we illustrate with vessel detection and flood-mapping tasks.
\end{abstract}

\input{sec/1_introduction}
\input{sec/2_related_works}
\input{sec/3_methodology}
\input{sec/4_results}

\input{sec/5_discussion}

\section*{LLM Usage Statement}
In this work we used Claude (Anthropic) Opus 5 and Sonnet 5 to assist in writing and refactoring plotting scripts, data-processing scripts and training code. We also used Opus 5 and GPT-5.6 (OpenAI) to assist in rewording and shortening parts of the text of our research paper draft. The research questions, methods, experiments, results and conclusions are the authors' own, and the authors are responsible for the whole content of this paper.

{
    \small
    \bibliographystyle{ieeenat_fullname}
    \bibliography{main}
}

\input{sec/appendix}


\end{document}

%% file: acronym.tex
\newacronym{ai}{AI}{artificial intelligence}
\newacronym{ml}{ML}{machine learning}
\newacronym{eo}{EO}{Earth observation}
\newacronym{esa}{ESA}{European Space Agency}

\newacronym[
    plural=GeoFMs,
    firstplural={geospatial foundation models (GeoFMs)}
]{geofm}{GeoFM}{geospatial foundation model}

\newacronym[
    plural=FMs,
    firstplural={foundation models (FMs)}
]{fm}{FM}{foundation model}

\newacronym{ssl}{SSL}{self-supervised learning}
\newacronym{mtl}{MTL}{multi-task learning}
\newacronym{da}{DA}{domain adaptation}
\newacronym{uda}{UDA}{unsupervised domain adaptation}
\newacronym{lulc}{LULC}{land use/land cover}
\newacronym{s2}{S2}{Sentinel-2}
\newacronym{gsd}{GSD}{ground sampling distance}

\newacronym{psd}{PSD}{power spectral density}

\newacronym{sar}{SAR}{synthetic aperture radar}
\newacronym{cr}{CR}{cognitive radar}
\newacronym{csar}{C-SAR}{cognitive synthetic aperture radar}
\newacronym{cfar}{CFAR}{constant false alarm rate}
\newacronym{prf}{PRF}{pulse repetition frequency}

\newacronym{rda}{RDA}{range--Doppler algorithm}
\newacronym{rcmc}{RCMC}{range cell migration correction}
\newacronym{csa}{CSA}{chirp scaling algorithm}
\newacronym{rma}{RMA}{range migration algorithm}
\newacronym{pga}{PGA}{phase gradient autofocus}
\newacronym{specan}{SPECAN}{spectral analysis}
\newacronym{fpga}{FPGA}{field programmable gate array}

\newacronym[
    plural=FFTs,
    firstplural={fast Fourier transforms (FFTs)}
]{fft}{FFT}{fast Fourier transform}

\newacronym{imu}{IMU}{inertial measurement unit}
\newacronym{gnss}{GNSS}{global navigation satellite system}

\newacronym[
    plural=CNNs,
    firstplural={convolutional neural networks (CNNs)}
]{cnn}{CNN}{convolutional neural network}

\newacronym{vit}{ViT}{Vision Transformer}
\newacronym{deit}{DeiT}{Data-efficient Image Transformer}
\newacronym{resnet}{ResNet}{residual network}
\newacronym{mim}{MiM}{masked image modeling}

\newacronym[
    plural=RNNs,
    firstplural={recurrent neural networks (RNNs)}
]{rnn}{RNN}{recurrent neural network}

\newacronym[
    plural=LSTMs,
    firstplural={long short-term memory networks (LSTMs)}
]{lstm}{LSTM}{long short-term memory network}

\newacronym[
    plural=GRUs,
    firstplural={gated recurrent units (GRUs)}
]{gru}{GRU}{gated recurrent unit}

\newacronym[
    plural=SSMs,
    firstplural={state space models (SSMs)}
]{ssm}{SSM}{state space model}

\newacronym{s4}{S4}{structured state space for sequence modeling}

\newacronym{s5}{S5}{simplified state space layer}

\newacronym{kd}{KD}{knowledge distillation}
\newacronym{crosskd}{CrossKD}{cross-architecture knowledge distillation}
\newacronym{at}{AT}{attention transfer}
\newacronym{rkd}{RKD}{relational knowledge distillation}
\newacronym{ckd}{CKD}{contrastive knowledge distillation}
\newacronym{mgd}{MGD}{masked generative distillation}
\newacronym{lora}{LoRA}{low-rank adaptation}
\newacronym{dr}{DR}{domain randomization}
\newacronym{mvg}{MVG}{multi-view generator}
\newacronym{dann}{DANN}{domain-adversarial neural network}
\newacronym{grl}{GRL}{gradient reversal layer}
\newacronym{adabn}{AdaBN}{adaptive batch normalization}
\newacronym{cycada}{CyCADA}{cycle-consistent adversarial domain adaptation}
\newacronym{mtor}{MTOR}{Mean Teacher with Object Relations}
\newacronym{umt}{UMT}{Unbiased Mean Teacher}

\newacronym{mmd}{MMD}{maximum mean discrepancy}
\newacronym{coral}{CORAL}{correlation alignment}
\newacronym{ot}{OT}{optimal transport}

\newacronym{gassl}{GASSL}{geography-aware self-supervised learning}
\newacronym{seco}{SeCo}{Seasonal Contrast}
\newacronym{satmae}{SatMAE}{Satellite Masked Autoencoder}
\newacronym{dofa}{DOFA}{Dynamic One-For-All}
\newacronym{croma}{CROMA}{Contrastive Radar-Optical Masked Autoencoder}
\newacronym{dino}{DINO}{Self-Distillation with No Labels}
\newacronym{moco}{MoCo}{Momentum Contrast}
\newacronym{simclr}{SimCLR}{Simple Framework for Contrastive Learning of Visual Representations}
\newacronym{ssl4eo}{SSL4EO}{self-supervised learning for Earth observation}

\newacronym{sift}{SIFT}{scale-invariant feature transform}
\newacronym{nn}{1-NN}{1-nearest neighbor}

\newacronym{mse}{MSE}{mean squared error}
\newacronym{rmse}{RMSE}{root mean squared error}
\newacronym{kl}{KL}{Kullback--Leibler divergence}
\newacronym{iou}{IoU}{intersection over union}
\newacronym{miou}{mIoU}{mean intersection over union}
\newacronym{pad}{PAD}{proxy A-distance}
\newacronym{psnr}{PSNR}{Peak Signal-to-Noise Ratio}

\newacronym{cpu}{CPU}{central processing unit}
\newacronym{gpu}{GPU}{graphics processing unit}
\newacronym{ram}{RAM}{random access memory}
\newacronym{vram}{VRAM}{video random access memory}
\newacronym{dsp}{DSP}{digital signal processing}
\newacronym{flops}{FLOPs}{floating-point operations}
\newacronym{otf}{OTF}{on-the-fly}

%% file: sec/1_introduction.tex
\section{Introduction}
\label{sec:intro}
\Gls{sar} is an active sensing modality that forms high-resolution images by transmitting microwave pulses from a moving platform, such as a plane or satellite, and recording the echoes reflected from the scene as the platform moves~\cite{curlander1991sar,cumming2005digital,richards2014sar} (Figure \ref{fig:intro-online-processing}). Rather than relying on a physically large antenna, \gls{sar} synthesizes a long aperture from the motion of the sensor, which enables fine azimuth resolution even from compact airborne or spaceborne platforms~\cite{curlander1991sar,carrara1995spotlight}. Because they operate at microwave frequencies, \gls{sar} systems can image the Earth during day and night and are substantially less affected by cloud cover or adverse weather than optical sensors~\cite{cumming2005digital,richards2014sar}.
\Gls{cr}, as introduced by Haykin~\cite{1593335}, is a concept that describes a perception--action loop. based on feedback from the observed environment~\cite{richards2014sar}, a radar adapts its parameters, such as beam steering, waveform, or the frequency of \gls{sar} pulses e.g. \gls{prf}.
Applying the closed-loop principle to \gls{csar} is compelling because \gls{sar} imaging is inherently subject to coverage--resolution trade-offs~\cite{curlander1991sar,cumming2005digital}: wide-swath modes enable rapid search, whereas narrow-swath/spotlight modes provide higher resolution for detailed inspection~\cite{carrara1995spotlight,jakowatz1996spotlight}.
However, in practice, closing the loop is challenging because raw \gls{sar} measurements must be focused before they become interpretable, and conventional focusing pipelines are computationally expensive and introduce substantial latency~\cite{cumming2005digital,soumekh1999sar,ulander2003fastbp}.
%\newline
In order to address these limitations and take a step towards a \gls{csar} system, this work makes the following contributions:
\begin{itemize}
  \item We re-formulate \gls{sar} image formation as online inference problem under the linear synthetic aperture assumption.
  \item We introduce StreamAZ, a compact state-space model for line-by-line azimuth focusing with fixed-size recurrent state, \mbox{\,$\sim$200\,} trainable parameters, and an explicit deployment mode that processes each incoming row without storing the full aperture.
  \item We develop an azimuth-focusing training loss that combines pointwise reconstruction with physics-constrained terms.
  \item We train the compact student model using knowledge distillation from a higher-capacity reference model obtained by multi-stage optimisation.
  \item We demonstrate a downstream proof-of-concept for low-latency decision support by applying \gls{cfar}-based vessel detection and flood segmentation in a line-by-line fashion.
\end{itemize}

%% file: sec/2_related_works.tex
\section{Related Work}

% Position relative to prior art; group works and state insufficiencies.
We position our contribution between (i) classical \gls{sar} image formation pipelines and (ii) modern sequence models for long-range operators.

\subsection{Image Formation}
\label{sec:related_sar_image_formation}

%In order to explain the novelty of our contribution, we first review how \gls{sar} focusing has traditionally been performed.
\gls{sar} systems are active sensors, meaning they use radar waves to illuminate a target and create an image based on the radar waves reflected from the target. However, this raw \gls{sar} data must undergo preprocessing in order to obtain an image with comprehensible structure;
in general, \gls{sar} image formation maps this raw complex phase history recorded by an antenna system to a reflectivity image by compensating for physical effects which have affected the received radar wave, such as range migration, azimuth phase evolution, motion errors, and system-dependent distortions.
%\todo[inline]{I think that the sentence about "what" sar specifically is isn't entirely clear to people without a SAR background}
Classical frequency-domain processors use \glspl{fft} in fast time and slow time (normally referring to data recorded as a sequence in the direction physically perpendicular to the motion of the satellite and parallel to the motion of the satellite respectively) to perform matched filtering on this raw data, with geometry-dependent phase correction and re-sampling. In this context, geometry-dependent phase correction refers to the need to mitigate for effects such as the changing distance of the \gls{sar} antenna to the illuminated target with respect to time.
%Classical frequency-domain processors perform matched filtering using \glspl{fft} in fast and slow time, with geometry-dependent phase corrections and re-sampling. 
The \gls{rda} separates focusing into range compression, \gls{rcmc}, and azimuth compression, and remains standard for stripmap data under moderate squint (the angular offset of the SAR radar antenna beam from the broadside/side-looking direction) and bandwidth assumptions~\cite{cumming2005digital,curlander1991sar,richards2014sar}.
Another commonly used focusing algorithm, the \gls{csa}, avoids explicit \gls{rcmc} interpolation through chirp scaling and phase multiplications~\cite{cumming2005digital,moreira1996csa}, while the so-called $\omega$--$k$ / \gls{rma} uses Stolt mapping, a computationally expensive interpolation step, to handle wide-angle spotlight and large-bandwidth data when its assumptions hold~\cite{carrara1995spotlight,soumekh1999sar,cumming2005digital,richards2014sar}. Polar-format and range-stack methods provide related spatial-frequency re-gridding strategies for spotlight data~\cite{jakowatz1996spotlight}. 
Time-domain backprojection directly evaluates the focusing integral by accumulating phase-corrected pulse contributions at each image location. It is robust to wide-angle collections, topography, and non-ideal trajectories, but computationally costly without factorized, subaperture, or parallel variants~\cite{carrara1995spotlight,soumekh1999sar,ulander2003fastbp}. 
Fast and subaperture methods can reduce latency and emit partial image updates, but they remain explicit aperture-integration algorithms whose memory and intermediate products depend on aperture support, scene geometry, and resolution.
Real-time and on-board \gls{sar} processing has largely focused on accelerating classical pipelines. \Gls{fpga}, \gls{gpu}, and embedded implementations of \gls{rda}, \gls{csa}, \gls{specan}, backprojection, and related methods use pipelined \glspl{fft}, reduced precision, hardware-aware memory layouts, and parallel scheduling to lower latency and downlink burden~\cite{parra2024onboard,choi2021fpga,mandapati2024floating,xu2022streamingbp,zhang2025embedded}. 
Notably, this notion of real time is mainly throughput-oriented and somewhat misleading: processing is accelerated in a real time manner \textit{only after data has been buffered along the time domain}, and this buffering makes the perception part of the perception-action loop of \gls{csar} too slow to be feasible. Because these accelerators rely on the traditional focusing algorithms described above, they can only begin focusing once the required aperture support is available, so this time-buffering is unavoidable.
Here, we instead ask whether azimuth compression itself can be approximated by a learned causal recurrence with aperture-independent state, enabling line-wise inference during acquisition. 
On the machine learning (ML) side, previous work has unrolled iterative
solvers into trainable networks~\cite{yonel2017passive, ji2024approxobs},
used deep image priors as regularisers~\cite{zhao2024deepred}, learned
end-to-end raw-to-image pipelines~\cite{rittenbach2020rdanet}, and learned
features from complex focused SAR data~\cite{huang2020deepsarnet}. These
methods operate offline on full echo arrays or focused products.
\subsection{Sequence models for physical 1D operators}
Learning-based components that operate on raw or partially processed radar signals are naturally framed as sequence-modeling problems.%, and here the main architectural families differ primarily in inductive bias and scaling behavior. 
Classical \glspl{rnn} and gated variants such as \glspl{lstm} and \glspl{gru} model sequences through hidden-state recursion with bounded memory~\cite{hochreiter1997lstm,cho2014gru}; however, their performance is limited on very-long sequences. Transformer models instead calculate the relationship between all tokens in a sequence directly via self-attention, making them more performative in long sequence tasks, however the computation and memory cost of self-attention scales poorly with the sequence length: in the canonical formulation the memory and computation cost scale as $\mathcal{O}(L^2)$ with the sequence length $L$~\cite{vaswani2017attention}; FlashAttention improves this memory cost to $\mathcal{O}(L)$ ~\cite{dao2022flashattention}. Modern \gls{ssm} models such as \gls{s4} represent a sequence through a latent dynamic state driven by the input and read out to the output. They are especially attractive for 1D signal processing because inference can be $\mathcal{O}(L)$ via a scan/recurrence and they have a fixed memory footprint of $\mathcal{O}(1)$ and perform well on long sequences.% Additionally, they have been shown to have high performance even when used on very long sequence tasks ~\cite{gu2022s4}.

%% file: sec/3_methodology.tex
\section{Methodology}
\label{sec:method}

%\subsection{SAR Acquisition Geometry}
%\label{sec:sar_acquisition_geometry}
Referring to Figure~\ref{fig:intro-online-processing}, we consider stripmap acquisition mode, in which the platform translates along the azimuth direction and the antenna beam is held fixed in azimuth relative to the platform, rather than being steered to dwell on a scene patch as in spotlight imaging.
Each received pulse provides samples in fast time, corresponding to range, and successive pulses sample slow time, corresponding to azimuth. Stacking the pulses therefore yields a two-dimensional complex phase-history array indexed by range bin and slow-time/pulse index.
Each pulse gives a range profile, and the same target is observed over many pulses as its sensor-target distance changes.
Because a target drifts across range bins during the aperture \cite{cumming2005digital}, 
classical processors explicitly compensate this effect through \gls{rcmc} or related re-sampling operations. In this work we don't compensate for this drift but we assume a moderate-aperture stripmap setting in which the synthetic aperture can be considered linear. This motivates the StreamAZ view of azimuth as an ordered slow-time sequence, where target information accumulates progressively across pulses.

\subsection{Online Azimuth Focuser}
Conceptually, our StreamAZ operates in three steps: (a) a complex range line is acquired, (b) range compression is applied using conventional \gls{dsp} operations, and (c) a compact neural state-space model incrementally performs azimuth focusing by updating a recurrent state that encodes the relevant phase-history context.
Instead of buffering a large azimuth aperture and applying explicit slow-time \glspl{fft}, the processor applies the same causal recurrent operator to each range cell in parallel. For range bin $r$ at pulse $k$, let $z_{k,r}\in\mathbb{R}^{2}$ denote the real/imaginary representation of the current complex sample. StreamAZ maintains a hidden state $s_{k,r}\in\mathbb{R}^{d}$ and updates it as
\begin{equation}
  s_{k+1,r} = g_{\theta}(s_{k,r}, z_{k,r}, p_r), \qquad \hat{x}_{k,r}=h_{\theta}(s_{k+1,r}),
  \label{eq:per-range-update}
\end{equation}
where $p_r$ is an optional normalized range-position feature and the parameters of $g_{\theta}$ and $h_{\theta}$ are shared across all range bins. This parameter sharing is what allows the model to focus an entire line while keeping the number of trainable weights extremely small.
In practice, $g_{\theta}$ is implemented as a lightweight input projection followed by a stack of S4D structured state-space layers and $h_{\theta}$ is a narrow output head that predicts the focused complex response. The deployable student contains only 208 trainable parameters, yet maintains enough dynamic state to model long azimuth responses. We initialize the recurrent kernels with the S4D-Lin parameterization~\cite{gu2022s4d}.
To stabilize the training of this very compact model, we adopt a distillation (\Cref{sec:teacher_student_distillation}) approach: first, a higher-capacity teacher is optimized for reconstruction quality under the azimuth-focusing objective described in \Cref{subsec:azimuth-loss}, and then the compact student is then trained to reproduce both the focused target and the teacher's response. 
During training we process finite azimuth strips in mini-batches using the convolutional mode of S4D, but at deployment we switch to the recurrent form of the model and retain only the current hidden state and the newest pulse. Reproducibility details are given in our supporting material appendix ~\ref{app:training_details}: our appendices are all available in our supporting material.

%% file: sec/4_results.tex
\section{Experiments}
\label{sec:experiments}
\label{sec:results}

We evaluate StreamAZ using a fixed Maya4 split, deterministic strip sampling, origin-preserving
complex normalization, and a recurrent deployment protocol matching the causal SSM
boundary conditions used during training.
\emph{Maya4} \cite{esa_philab_maya4_2025} is a Sentinel-1-derived dataset  of 2TB, of which we split 300GB into training, validations and test data, that provides intermediate \gls{sar} representations from Level~0 raw measurements to Level 1 focused imagery. Using the \gls{rda} as the reference pipeline, each sample is stored as four aligned stages: \emph{raw}, \emph{range-compressed} (RC), \emph{range-cell-migration-corrected} (RCMC), and \emph{azimuth-compressed} (AZ).

\subsection{Azimuth Focusing Loss}
\label{subsec:azimuth-loss}
The azimuth-focusing loss is a central contribution of this study. We supervise azimuth focusing through three complementary components, namely reconstruction fidelity ($\mathcal{L}_{\mathrm{rec}}$), distribution matching ($\mathcal{L}_{\mathrm{dist}}$), and
structural consistency ($\mathcal{L}_{\mathrm{struct}}$).
Let $\hat{z},z\in\mathbb{C}^{L}$ denote the predicted and reference
azimuth strips after inverse min--max normalization.
We define $a=|\hat{z}|$, $a^\star=|z|$,
$\ell=\log(a+\varepsilon)$, and
$\ell^\star=\log(a^\star+\varepsilon)$, where $\varepsilon>0$.
The overall objective is
\begin{equation}
\label{eq:az_focus_loss}
    \mathcal{L}_{\mathrm{AF}}
    =
    \mathcal{L}_{\mathrm{rec}}
    +
    \mathcal{L}_{\mathrm{dist}}
    +
    \mathcal{L}_{\mathrm{struct}}.
\end{equation}
Each component is a weighted non-negative sum of the subterms we define in
Table~\ref{tab:loss_function}, and further elaborate in Appendix A.
Here, $\rho(\cdot,\cdot)$ denotes the amplitude-correlation
coefficient, $Q_p(\cdot)$ the empirical $p$-quantile, and
$\Delta$ and $\Delta^2$ the first- and second-order finite
differences along azimuth.
$P(\cdot)$ denotes the normalized one-dimensional azimuth \gls{psd} estimate evaluated over normalized frequencies
$f\in[0.15,0.5]$, and $W(\cdot)$ is an autocorrelation-derived
focus-width statistic.

\begin{table}[b]
  \centering
  \caption{Subterms and weights of the azimuth-focusing
    loss $\mathcal{L}_{\mathrm{AF}}$.}
  \label{tab:loss_function}
  \footnotesize
  \setlength{\tabcolsep}{22pt}
  \renewcommand{\arraystretch}{1.2}
  \begin{tabular}{@{}llc@{}}
    \toprule
    \textbf{Term} & \textbf{Definition} & \textbf{Weight} \\
    \midrule

    $\mathcal{L}_{\mathrm{complex}}$ &
    $\frac{1}{L}\|\hat z-z\|_1$ &
    $w_c$ \\

    $\mathcal{L}_{\log\mathrm{amp}}$ &
    $\frac{1}{L}\|\ell-\ell^\star\|_1$ &
    $w_{\log}$ \\

    $\mathcal{L}_{\mathrm{amp\text{-}corr}}$ &
    $1-\rho(a,a^\star)$ &
    $w_{\mathrm{ac}}$ \\

    $\mathcal{L}_{\mathrm{tail}}$ &
    $\begin{aligned}
      &\left|\log\frac{Q_{.95}(a)}{Q_{.95}(a^\star)}\right|\\
      &+\frac12\left|
      \log\frac{Q_{.99}(a)}{Q_{.99}(a^\star)}
      \right|
    \end{aligned}$ &
    $w_{\mathrm{tail}}$ \\

    $\mathcal{L}_{\mathrm{az\text{-}grad}}$ &
    $\frac{\|\Delta\ell-\Delta\ell^\star\|_1}{L-1}$ &
    $w_{\nabla}$ \\

    $\mathcal{L}_{\mathrm{az\text{-}psd}}$ &
    $\|P(\ell)-P(\ell^\star)\|_1$ &
    $w_{\mathrm{psd}}$ \\

    $\mathcal{L}_{\mathrm{focus\text{-}width}}$ &
    $\frac{|W(\ell)-W(\ell^\star)|}
      {W(\ell^\star)+\varepsilon}$ &
    $w_{\mathrm{fw}}$ \\

    $\mathcal{L}_{\mathrm{az\text{-}edge}}$ &
    $\frac{\|\Delta^2\ell-\Delta^2\ell^\star\|_1}{L-2}$ &
    $w_{\mathrm{edge}}$ \\

    \bottomrule
  \end{tabular}
\end{table}

% \begin{table*}[t]
%   \centering
%   \caption{Subterms and weights of the azimuth-focusing loss
%     $\mathcal{L}_{\mathrm{AF}}$.}
%   \label{tab:loss_function}
%   \footnotesize
%   \setlength{\tabcolsep}{6pt}
%   \renewcommand{\arraystretch}{1.15}
%   \begin{tabular}{@{}llc@{\qquad}llc@{}}
%     \toprule
%     \textbf{Term} & \textbf{Definition} & \textbf{Weight}
%     & \textbf{Term} & \textbf{Definition} & \textbf{Weight} \\
%     \midrule

%     $\mathcal{L}_{\mathrm{complex}}$ &
%     $\tfrac{1}{L}\lVert \hat{z}-z\rVert_1$ &
%     $w_c$ &
%     $\mathcal{L}_{\log\mathrm{amp}}$ &
%     $\tfrac{1}{L}\lVert \ell-\ell^\star\rVert_1$ &
%     $w_{\log}$ \\

%     $\mathcal{L}_{\mathrm{amp\text{-}corr}}$ &
%     $1-\rho(a,a^\star)$ &
%     $w_{\mathrm{ac}}$ &
%     $\mathcal{L}_{\mathrm{tail}}$ &
%     $\bigl|\!\log\tfrac{Q_{.95}(a)}{Q_{.95}(a^\star)}\bigr|
%       +\tfrac{1}{2}
%        \bigl|\!\log\tfrac{Q_{.99}(a)}{Q_{.99}(a^\star)}\bigr|$ &
%     $w_{\mathrm{tail}}$ \\

%     $\mathcal{L}_{\mathrm{az\text{-}grad}}$ &
%     $\tfrac{1}{L\!-\!1}
%       \lVert \Delta\ell-\Delta\ell^\star\rVert_1$ &
%     $w_{\nabla}$ &
%     $\mathcal{L}_{\mathrm{az\text{-}psd}}$ &
%     $\lVert P(\ell)-P(\ell^\star)\rVert_1$ &
%     $w_{\mathrm{psd}}$ \\

%     $\mathcal{L}_{\mathrm{focus\text{-}width}}$ &
%     $\frac{|W(\ell)\!-\!W(\ell^\star)|}
%       {W(\ell^\star)+\varepsilon}$ &
%     $w_{\mathrm{fw}}$ &
%     $\mathcal{L}_{\mathrm{az\text{-}edge}}$ &
%     $\tfrac{1}{L\!-\!2}
%       \lVert \Delta^2\ell-\Delta^2\ell^\star\rVert_1$ &
%     $w_{\mathrm{edge}}$ \\

%     \bottomrule
%   \end{tabular}
% \end{table*}

\paragraph{Reconstruction.}
$\mathcal{L}_{\mathrm{rec}}$ combines complex-field reconstruction
and log-amplitude matching:
\begin{equation}
\label{eq:az_focus_rec}
    \mathcal{L}_{\mathrm{rec}}
    =
    w_c\,\mathcal{L}_{\mathrm{complex}}
    +
    w_{\log}\,\mathcal{L}_{\log\mathrm{amp}}.
\end{equation}

The complex term $\mathcal{L}_{\mathrm{complex}}$ is the primary data-fidelity anchor. It acts in complex space, penalising both residual phase and amplitude mismatch. 
The log-amplitude term $\mathcal{L}_{\log\mathrm{amp}}$ compresses the dynamic range because its purpose is to reduce the influence of very strong scatterers (completely out-of-distribution). In simple words this makes multiplicative amplitude
errors approximately additive when amplitudes dominate $\varepsilon$.

\paragraph{Distribution.}
$\mathcal{L}_{\mathrm{dist}}$ combines amplitude correlation and
upper-tail matching:
\begin{equation}
\label{eq:az_focus_stat}
    \mathcal{L}_{\mathrm{dist}}
    =
    w_{\mathrm{ac}}\,\mathcal{L}_{\mathrm{amp\text{-}corr}}
    +
    w_{\mathrm{tail}}\,\mathcal{L}_{\mathrm{tail}}.
\end{equation}
The correlation term emphasizes agreement in the azimuthal amplitude
pattern rather than absolute gain.
The tail term compares the $95$th and $99$th amplitude quantiles,
providing explicit supervision of high-intensity returns beyond
average reconstruction error.
These terms therefore compare both the relative amplitude pattern
and the upper-tail statistics.
\paragraph{Structure.}
$\mathcal{L}_{\mathrm{struct}}$ compares local and aggregate structure
along the log-amplitude profiles:
\begin{equation}
\label{eq:az_focus_structure}
\begin{aligned}
\mathcal{L}_{\mathrm{struct}}
={}& w_{\nabla}\,\mathcal{L}_{\mathrm{az\text{-}grad}}
  + w_{\mathrm{psd}}\,\mathcal{L}_{\mathrm{az\text{-}psd}} \\
 &{}+ w_{\mathrm{fw}}\,\mathcal{L}_{\mathrm{focus\text{-}width}}
  + w_{\mathrm{edge}}\,\mathcal{L}_{\mathrm{az\text{-}edge}}.
\end{aligned}
\end{equation}
The gradient and edge terms match first- and second-order finite
differences, respectively, penalizing discrepancies in transitions
and local curvature.
The \gls{psd} term compares spectral content within the selected azimuth
frequency band.
The focus-width term penalizes deviations from the reference
autocorrelation-derived width, including both broadening and
narrowing.
Together, these terms complement pointwise reconstruction with
structural constraints relevant to azimuth focusing.

\input{sec/opt_in_multi_stage_teacher_optimisation}

\subsection{Teacher--Student Distillation}
\label{sec:teacher_student_distillation}

%%%%%%%%%%%%%%%%%%%%%%%%%%%%%%% TABLE
\input{tables/teacher_student_comparison}
%%%%%%%%%%%%%%%%%%%%%%%%%%%%%%% TABLE

The staged sweep selected a high-capacity teacher $T_{\phi}$, trained with the
azimuth-focusing loss and used only for distillation. The deployable streamAZ is the compact student $S_{\theta}$, which keeps the same complex I/O
interface but uses the recurrent state-space stack needed for linewise online inference.
Distillation is used as both a compression
mechanism and an optimization stabilizer: the student is not trained to discover the full azimuth-compression operator from the hard AZ target alone, but is also given a
smooth complex-valued surrogate response from the teacher.
\begin{figure}[htb]
  \centering
  \includegraphics[
    width=0.75 \linewidth,
    trim=0.cm 0cm 0cm 0cm,
    clip
  ]{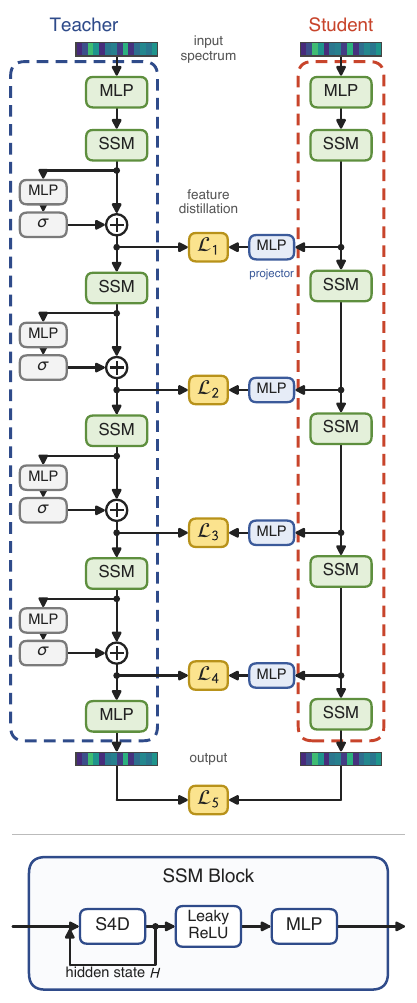}
  \caption{
  Teacher--student distillation architecture for StreamAZ.  The offline teacher uses a
  higher-capacity SSM stack to learn a stable complex-valued RC-to-AZ focusing
  map under the azimuth-focusing objective.  
  %During distillation, the teacher is
  %frozen and provides soft complex-field supervision to the compact student. 
  The
  student preserves the same input--output interface as the teacher, but uses a
  much smaller causal SSM stack that can be executed in recurrent mode for
  fixed-state online inference.
  }
  \label{fig:distillation-architecture}
\end{figure}

\begin{figure*}[t]
\centering
\includegraphics[
    width=\linewidth]{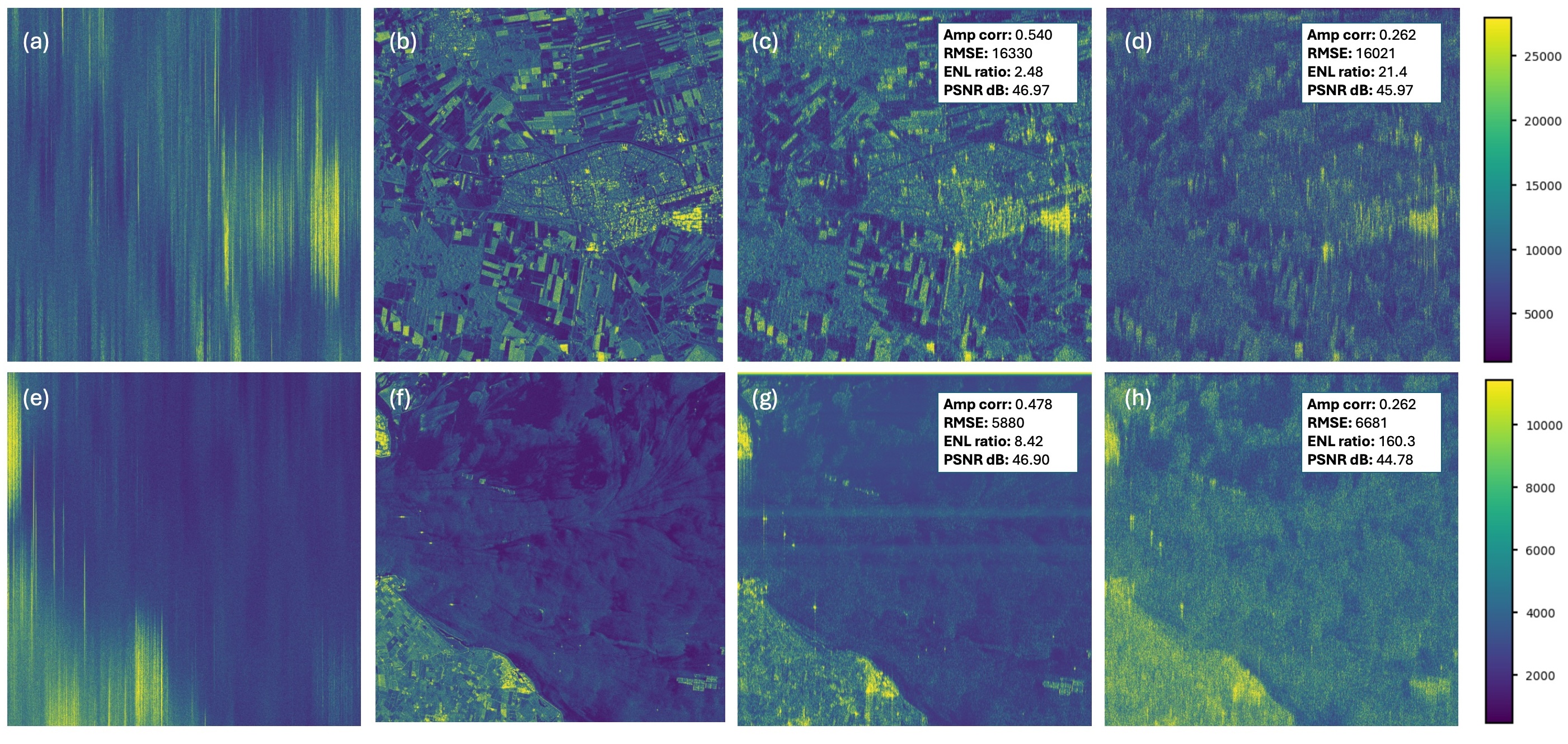}
\caption{
Teacher--student inference on two held-out Sentinel-1-derived SAR strips.
The top row corresponds to product 0042df and the bottom row to product 00413d.
Panels (a,e) are
the RC inputs, (b,f) are the AZ references (produced using the traditional \gls{rda} algorithm, (c,g) are teacher outputs, and (d,h)
are student outputs. Metrics overlaid on the prediction panels are computed
against the AZ reference: amplitude correlation, RMSE, equivalent-number-of-looks
ratio (used to quantify the amount of speckle-related noise in a SAR image), and \gls{psnr}. %The teacher more closely preserves the reference scene
morphology and radiometric structure, while the compact student recovers the
coarse focused layout but exhibits lower amplitude correlation and stronger
strip-wise artifacts.
}
\label{fig:teacher_student_qualitative}
\end{figure*}

\begin{figure*}[t]
  \centering
  \begin{minipage}{0.8\textwidth} % Adjust this percentage (e.g., 0.8 to 0.9) to control total width
    \centering
    \includegraphics[width=0.31\textwidth, trim=0cm 3.6cm 0cm 0cm, clip]{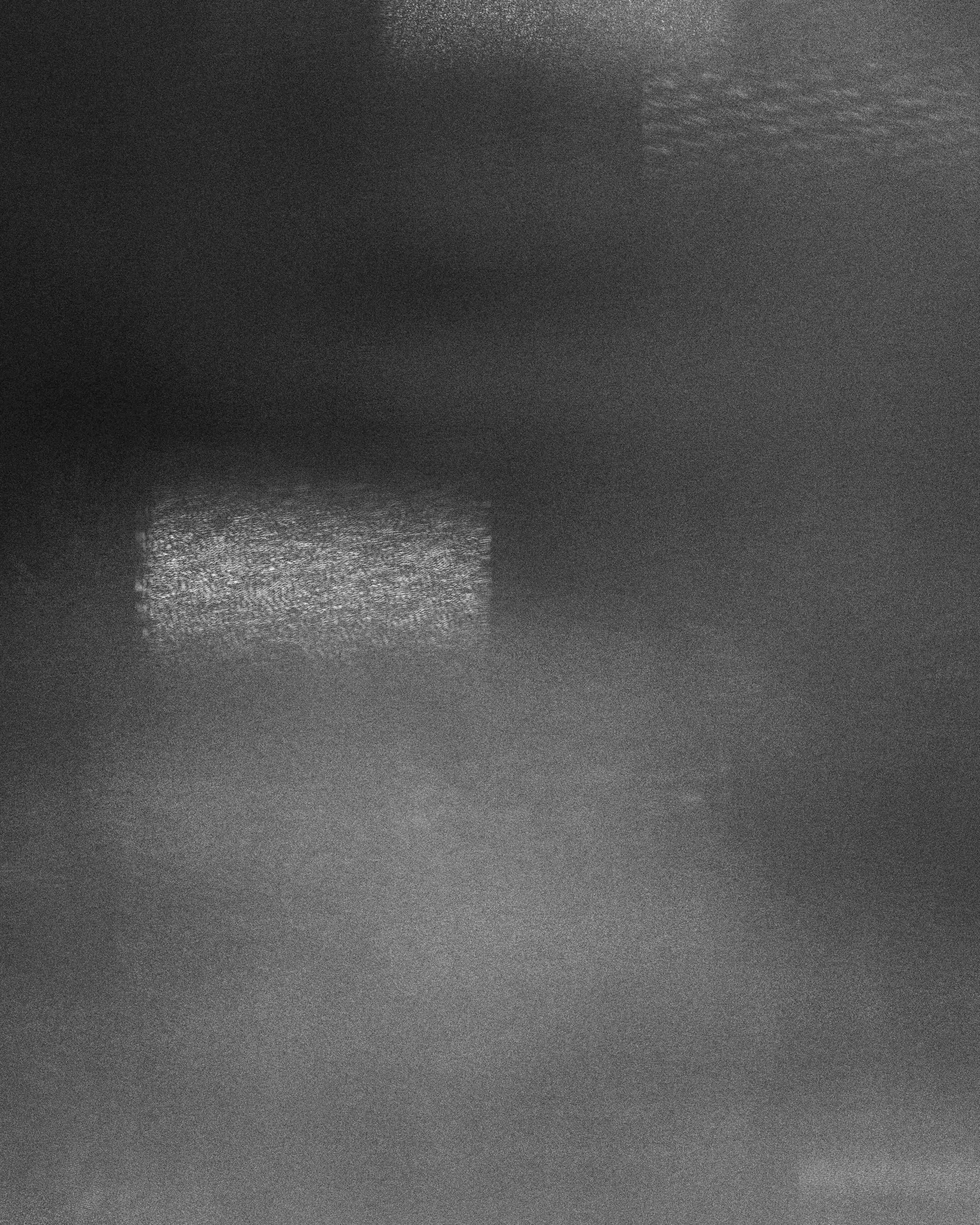}
    \hfill
    \includegraphics[width=0.31\textwidth, trim=0cm 1.2cm 0cm 0cm, clip]{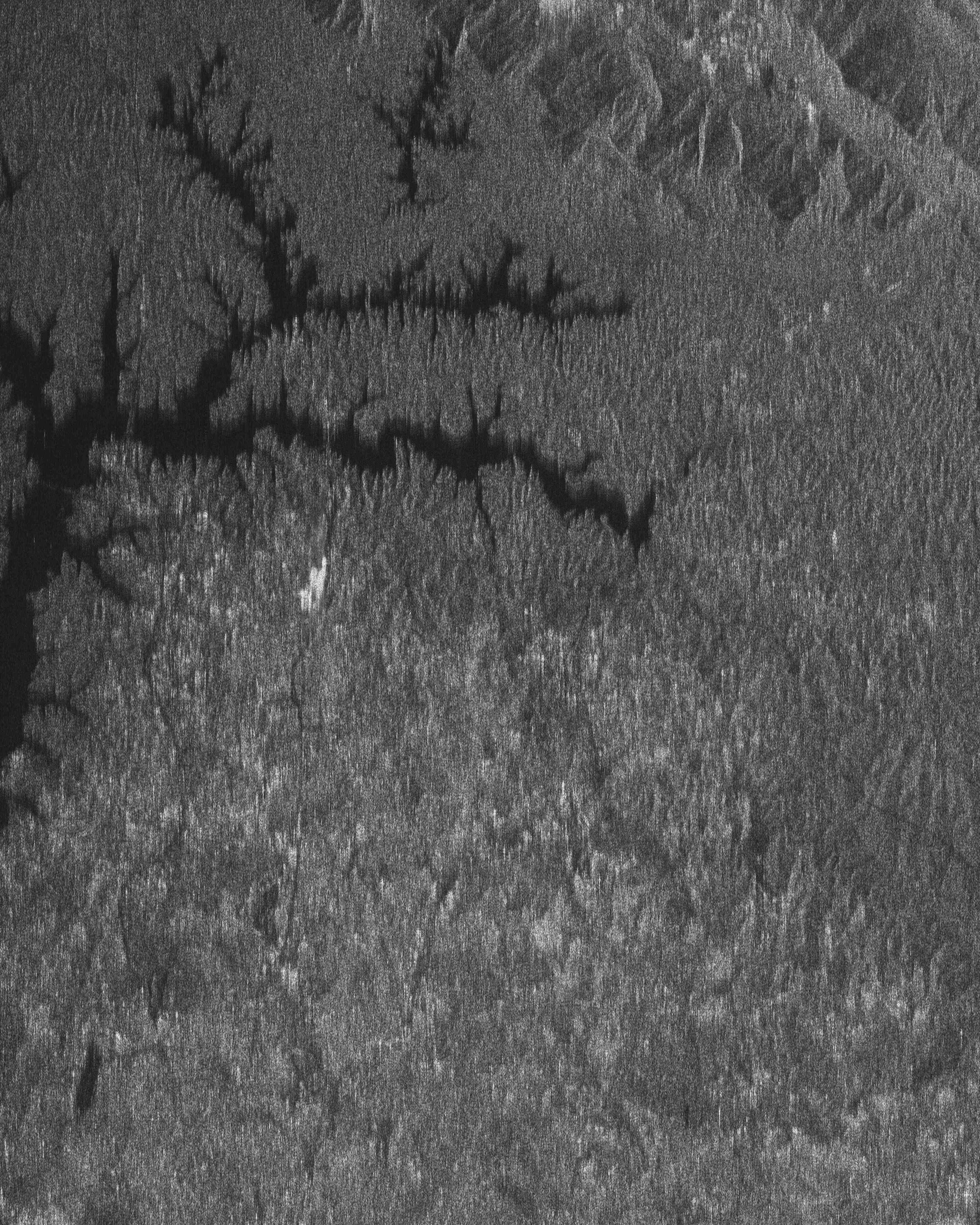}
    \hfill
    \includegraphics[width=0.31\textwidth, trim=0cm 3.6cm 0cm 0cm, clip]{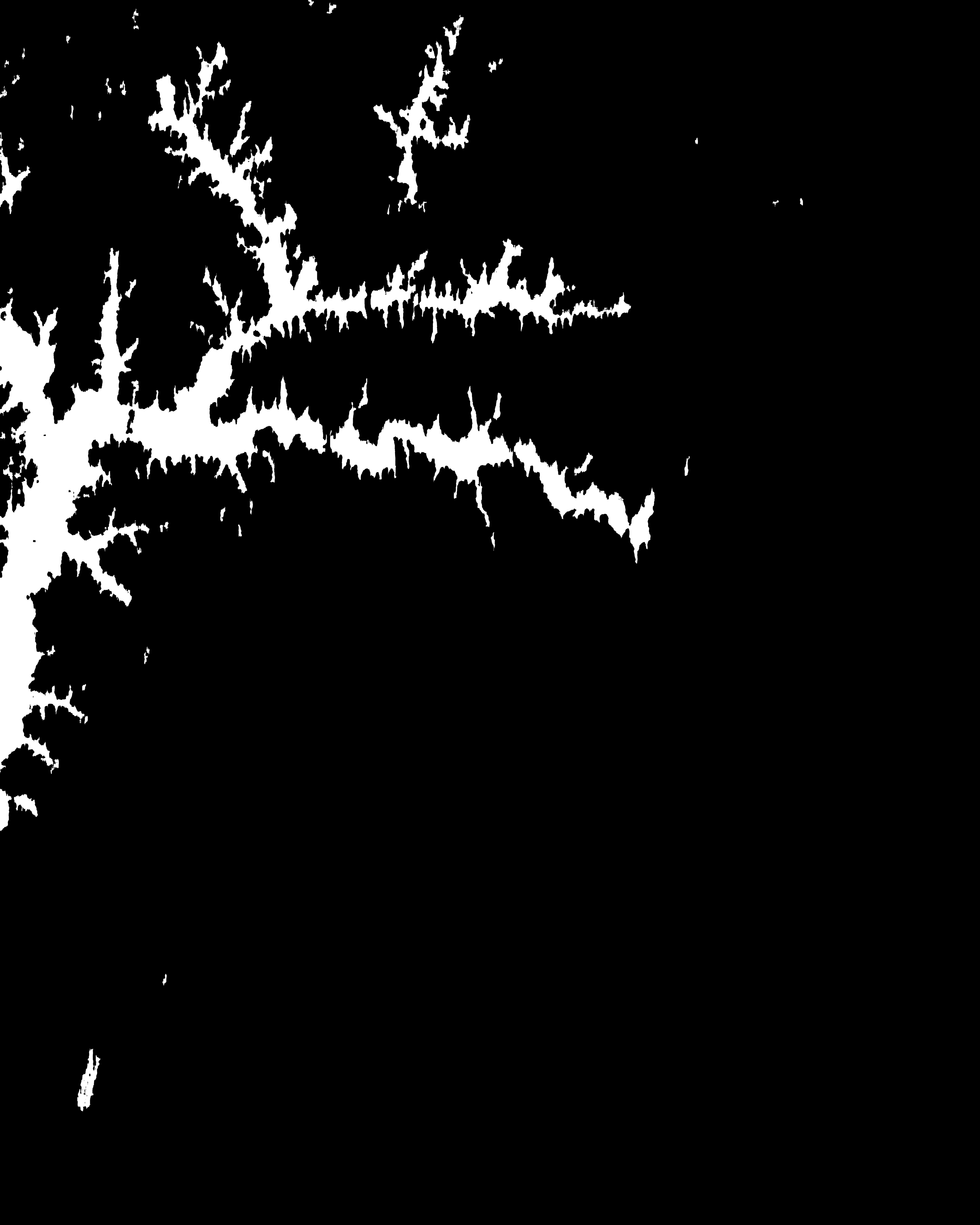}
    
    \vspace{6pt} % vertical spacing between rows
    
    \includegraphics[width=0.31\textwidth]{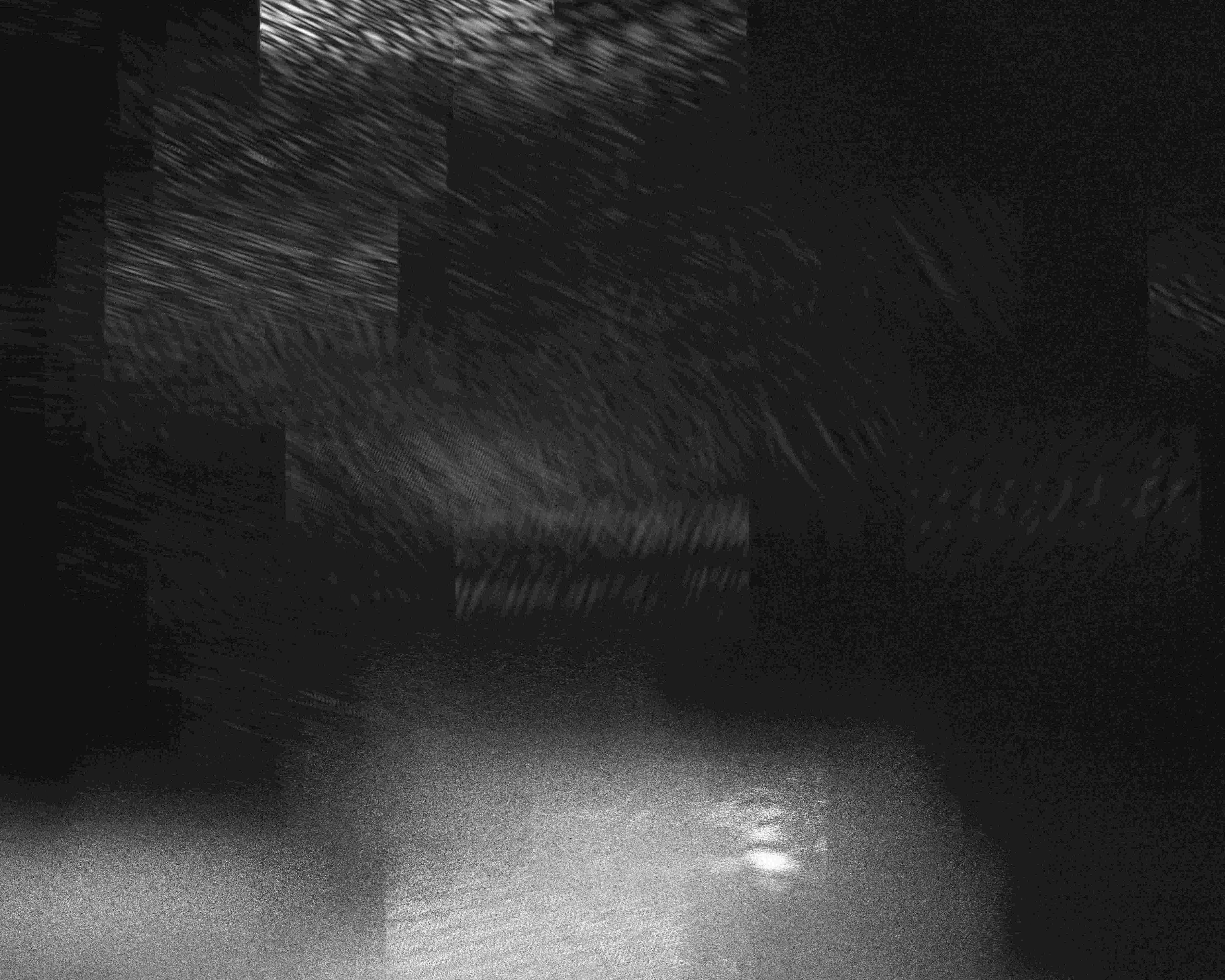}
    \hfill
    \includegraphics[width=0.31\textwidth]{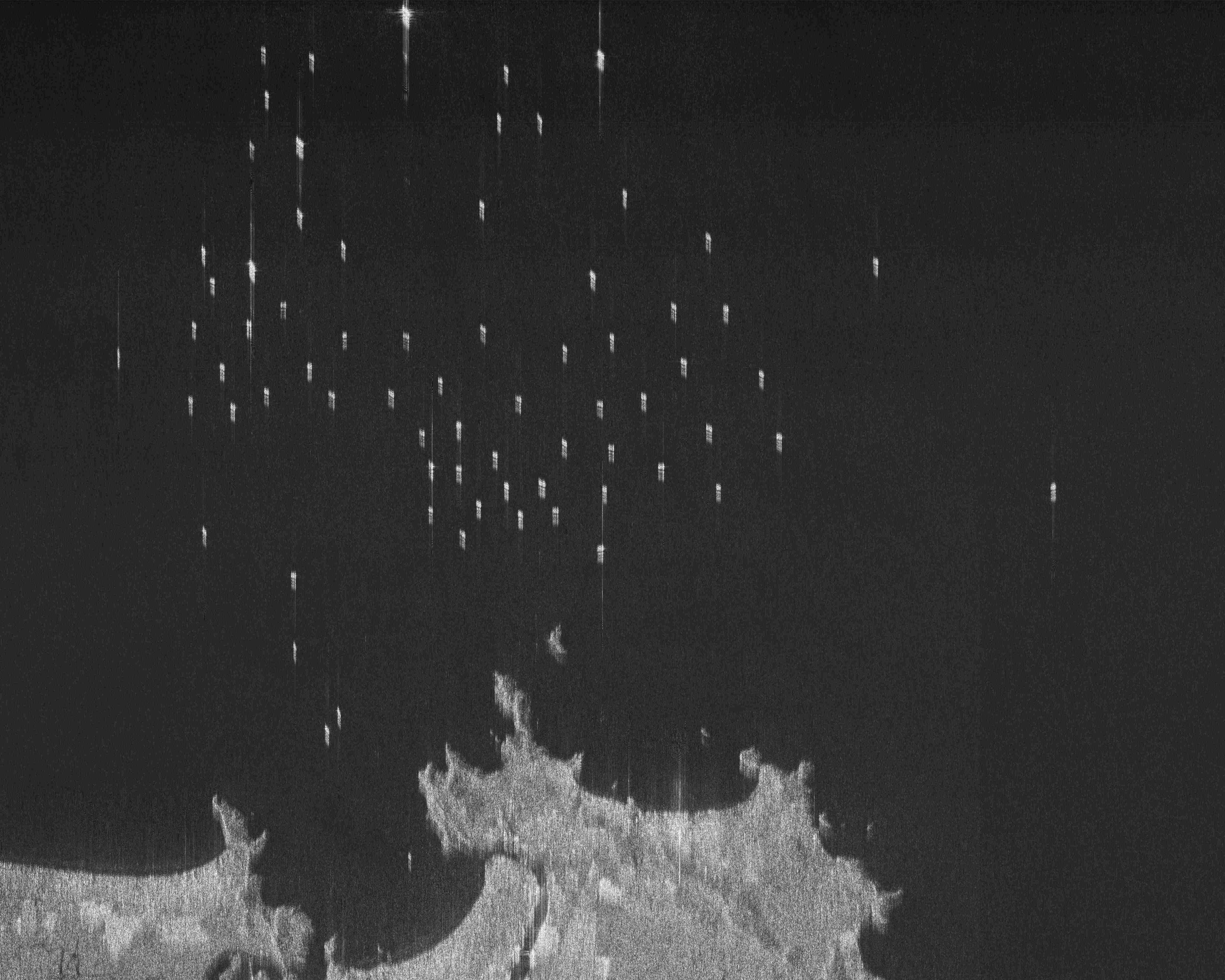}
    \hfill
    \includegraphics[width=0.31\textwidth]{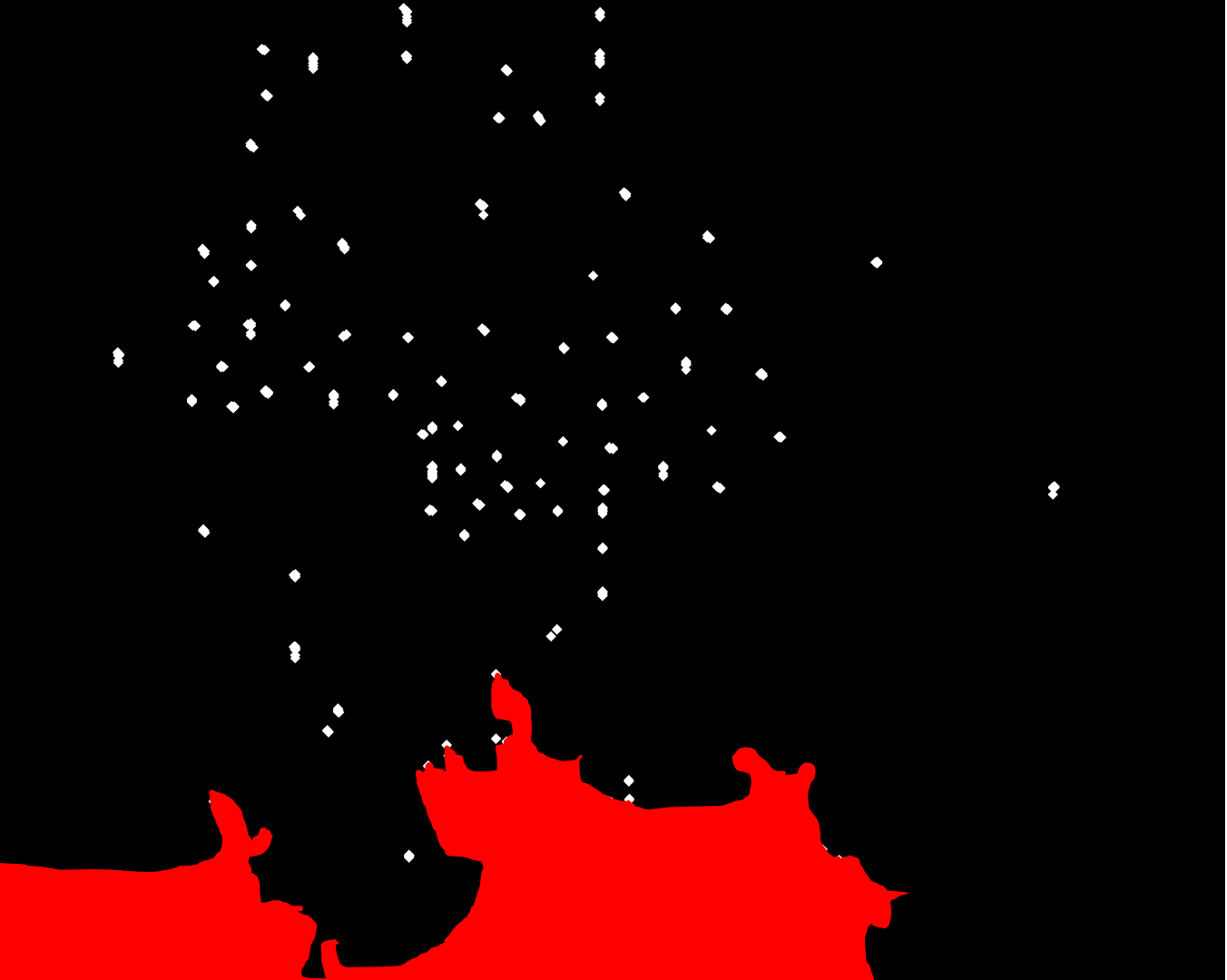}
  \end{minipage}
  
  \caption{
  (Top) Demonstration of constant threshold water segmentation with applications such as flood detection - scan captured over Sao Paolo. (Top Left): raw SAR data as received by satellite. (Top Center): focused output produced by StreamAZ. (Top Right): Segmentation map of water coverage obtained using single value thresholding. (Bottom) Demonstration of vessel detection using StreamAZ over Porto de Santos, Brazil.
  (Bottom Left): raw SAR data as received by the satellite.
  (Bottom Center): focused output produced by StreamAZ; bright point responses correspond to strong scatterers, including vessels and port infrastructure.
  (Bottom Right): vessel-like detections obtained with a linewise CFAR detector using a rolling buffer of 77 range lines. False positives over land are suppressed using geographic masking, shown in red.
  }
  \label{fig:ship_detection}
\end{figure*}

For an inverse-normalized input azimuth strip $u$, reference AZ strip $z$, teacher
prediction $\hat z_T=T_{\phi}(u)$, and student prediction
$\hat z_S=S_{\theta}(u)$, the teacher parameters are frozen and the student is trained
with
\begin{equation}
\mathcal{L}_{\mathrm{student}}
= \mathcal{L}_{\mathrm{AF}}(\hat z_S,z)
+ \lambda_{\mathrm{KD}}\,
\mathcal{D}\!
\left(\hat z_S,\operatorname{sg}[\hat z_T]\right),
\label{eq:student_distillation_loss}
\end{equation}
where $\operatorname{sg}[\cdot]$ denotes stop-gradient and
$\mathcal{L}_{\mathrm{AF}}$ is the physically motivated azimuth-focusing loss defined
in Eq.~\ref{eq:az_focus_loss}.  
With reference to Figure \ref{fig:distillation-architecture}, the distillation discrepancy is evaluated in the same complex domain as
the reconstruction loss,
\begin{equation}
\begin{split}
\mathcal{D}(a,b) ={}&
\alpha_c \|a-b\|_1
+ \alpha_{\ell}\left\|\log(|a|+\epsilon)-\log(|b|+\epsilon)\right\|_1 \\
&+ \alpha_{\rho}\left(1-\rho(|a|,|b|)\right)
+ \alpha_{\phi}\left\langle 1-\cos\left(\angle a-\angle b\right)\right\rangle .
\end{split}
\label{eq:complex_distillation_discrepancy}
\end{equation}
The first term transfers the teacher's complex response, the log-amplitude and
correlation terms transfer radiometric structure while being less sensitive to global
scale, and the circular phase term avoids an artificial penalty at the $-\pi/\pi$
wrap.  The hard-target term in Eq.~\eqref{eq:student_distillation_loss} prevents the
student from inheriting teacher-specific bias, while the teacher term regularizes the
severely under-parameterized recurrent model.  After training, $T_{\phi}$ is discarded; only $S_{\theta}$ is used in the
online processor.
Table~\ref{tab:teacher_student_distillation} summarizes the quantitative
teacher--student gap on the held-out set. 
The distilled student largely preserves
aggregate pointwise accuracy: RMSE increases by only $2.16\%$, while phase MAE and
phase RMSE increase by $11.2\%$ and $8.51\%$, respectively. 
The larger degradation is instead in structured complex fidelity. 
The representative qualitative results in
\Cref{fig:teacher_student_qualitative} explain the aggregate teacher--student
gap.  The teacher predictions preserve the dominant morphology of both held-out
scenes: the agricultural field pattern in product 0042df and the coastal/water
structure in product 00413d remain spatially aligned with the AZ references.  This
is reflected in the overlaid amplitude-correlation scores, where the teacher
outperforms the student in both examples (\(0.540\) versus \(0.262\) for 0042df,
and \(0.478\) versus \(0.262\) for 00413d).  The student recovers the coarse
focused layout, but its output is less structurally faithful.
Thus, the student model is good enough for rough amplitude-based applications but is not yet reliable for tasks that need accurate phase or calibrated complex outputs.

\subsection{Potential Online  Applications}

An important test of SAR focusing quality is whether the focused output supports simple downstream decisions. We therefore experiment with two (of many) potential downstream tasks that would benefit from rapid focusing using StreamAZ; these two applications are CFAR vessel detection and fixed threshold flood mapping, the results of which we show in figure~\ref{fig:ship_detection}. Vessel detection is important for preservation of protected maritime areas where fishing and shipping vessel passage are restricted. By providing a rapid localization of a vessel in a restricted maritime area using StreamAZ and linewise CFAR, it could be possible for the SAR system to take a second, high-resolution scan via spotlight whilst still flying overhead the detected vessel of interest for identification purposes and improve enforcement on fishing restrictions etc. As a proof-of-concept we apply a linewise CFAR detector to an StreamAZ-focused scene over Porto de Santos, Brazil. Rapid flood-mapping is also another time-sensitive downstream task which would benefit from rapid focusing to support early situational awareness and sensor-tasking decisions. As an example, we implement a fixed-threshold based segmentation of a StreamAZ-focused scan of lakes in the area around Sao Paolo, also presented in figure~\ref{fig:ship_detection}. Our intent is to demonstrate that many applications would benefit from rapid focusing for re-tasking and that the output quality of StreamAZ is high enough to achieve this.

%% file: sec/opt_in_multi_stage_teacher_optimisation.tex
\subsection{Multi-Stage Teacher Optimization}
\label{subsec:teacher-optimization}

\begin{table}[b]
  \centering
  \begin{minipage}{\columnwidth}
    \footnotesize
    \setlength{\tabcolsep}{2pt}
    \renewcommand{\arraystretch}{1.08}

    % CVPR's warning tolerance also affects tabularx width fitting.
    \makeatletter
    \TX@delta=0.1pt
    \makeatother

    \caption{Multi-stage teacher optimization.}
    \label{tab:teacher-optimization-unified}

    \captionsetup[subtable]{
      font=footnotesize,
      labelfont=bf,
      textfont=bf,
      justification=raggedright,
      singlelinecheck=false,
      skip=2pt
    }

    \begin{subtable}{\linewidth}
      \caption{Stage representatives; \(N\) denotes the number of runs.}
      \label{tab:teacher-performance}

      \begin{tabularx}{\linewidth}{
        @{}l*{6}{>{\raggedleft\arraybackslash}X}@{}
      }
        \toprule
        Stage &
        \(N\) &
        \(R\downarrow\) &
        \(C\uparrow\) &
        \(Q\uparrow\) &
        \(\Delta_R\) &
        \(\Delta_C\) \\
        \midrule

        NA\(^{\dagger}\) &
        16 &
        \multicolumn{5}{c}{Failed to converge} \\

        \(0^{*}\) &
        1 &
        7837.360 &
        0.2063 &
        0.2806 &
        \(0.00\%\) &
        \(0.00\%\) \\
        \midrule

        I &
        9 &
        7747.460 &
        0.213 &
        0.291 &
        \(+1.15\%\) &
        \(+3.10\%\) \\

        II &
        27 &
        \textbf{7745.140} &
        0.215 &
        0.299 &
        \(\mathbf{+1.18\%}\) &
        \(+4.36\%\) \\

        III &
        6 &
        7816.940 &
        0.207 &
        0.282 &
        \(+0.26\%\) &
        \(+0.48\%\) \\

        IV &
        16 &
        7808.830 &
        0.217 &
        0.298 &
        \(+0.37\%\) &
        \(+5.28\%\) \\

        \rowcolor{gray!12}
        V &
        92 &
        7752.610 &
        \textbf{0.222} &
        \textbf{0.303} &
        \(+1.08\%\) &
        \(\mathbf{+7.51\%}\) \\

        \bottomrule
      \end{tabularx}
    \end{subtable}

    \par\smallskip

    \begin{subtable}{\linewidth}
      \caption{Loss settings: Stage~0 \(\to\) teacher V.}
      \label{tab:teacher-loss-settings}

      \begin{tabularx}{\linewidth}{
        @{}*{3}{l>{\raggedleft\arraybackslash}X}@{}
      }
        \(w_c\) &
        \(\text{---}\to0.04\) &
        \(w_{\log}\) &
        \(0.22\to0.18\) &
        \(w_{\mathrm{ac}}\) &
        \(0.20\to0.18\) \\

        \(w_{\mathrm{tail}}\) &
        \(0.14\to0.12\) &
        \(w_{\nabla}\) &
        \(0.12\to0.24\) &
        \(w_{\mathrm{psd}}\) &
        \(0.08\to0.04\) \\

        \(w_{\mathrm{fw}}\) &
        \(0.02\to0.04\) &
        \(w_{\mathrm{edge}}\) &
        \(0.00\to0.08\) &
        \(\epsilon\) &
        \(\text{---}\to10^{-3}\) \\

        \bottomrule
      \end{tabularx}
    \end{subtable}

    \par\smallskip\noindent
    \(^{*}\)Development StreamAZ baseline; no sweep.
    \(\eta=4.0\times10^{-4}\),
    \(d_{\mathrm{state}}=512\),
    \(L=4\),
    dropout \(=0.05\),
    selectivity off, skip on.
    Dashes indicate values not specified for Stage~0.

    \par\smallskip\noindent
    \(\Delta_R=100(R_0-R)/R_0\),
    \(\Delta_C=100(C-C_0)/C_0\).
    Larger values indicate lower RMSE and higher amplitude correlation,
    respectively.

    \par\smallskip\noindent
    \(^{\dagger}\)Debug grid: LSTM/GRU; \(h\): hidden dimension;
    \(h\in\{6,9\}\), \(L\in\{1,2\}\),
    \(\eta\in\{10^{-4},3\times10^{-4}\}\).
    All 16 runs used input/output dimensions \(3/1\),
    unidirectional recurrence and dropout \(0.0\).
    No teacher was selected.
    Separate production baselines used GRU \(h=11,L=4\)
    and LSTM \(h=9,L=4\), both with dropout \(0.05\).
    \par
  \end{minipage}
\end{table}
To tune our model, we perform staged grid searches around a shared baseline (Stage-0). \Cref{tab:teacher-performance} reports the Stage-0 baseline and the selected
run at each stage. All improvements are computed relative to Stage~0.
The baseline and final-teacher loss settings are compared in
\Cref{tab:teacher-loss-settings}. All completed runs were evaluated using
RMSE \(R\), amplitude correlation \(C\), and complex coherence \(Q\).
Search spaces and selected changes are detailed below.
The setting selected at each stage minimizes the rank score
\begin{equation}
  s(x)
  =
  \frac{1}{2}
  \Bigl(
    \operatorname{rk}_{R}(x)
    +
    \operatorname{rk}_{C}(x)
  \Bigr),
  \label{eq:teacher-rank-score}
\end{equation}
where \(\operatorname{rk}_{R}\) is the dense rank of RMSE in ascending order
and \(\operatorname{rk}_{C}\) is the dense rank of amplitude correlation
in descending order. Complex coherence is reported as an auxiliary
diagnostic and is not used directly in the rank score.
Phase preservation for phase-critical products, such as interferometry,
is left to future work.
When we refer to ``selectivity'' in the ablations, we mean a lightweight
input-conditioned gate in our streamAZ block, not the full selective-scan
architecture introduced by Mamba \cite{gu2023mamba}.
The gate is used only to test whether modest input-dependent modulation
of the state update improves the azimuth-focusing surrogate.
\begin{figure}[t]
  \centering
  \includegraphics[width=0.9\linewidth]{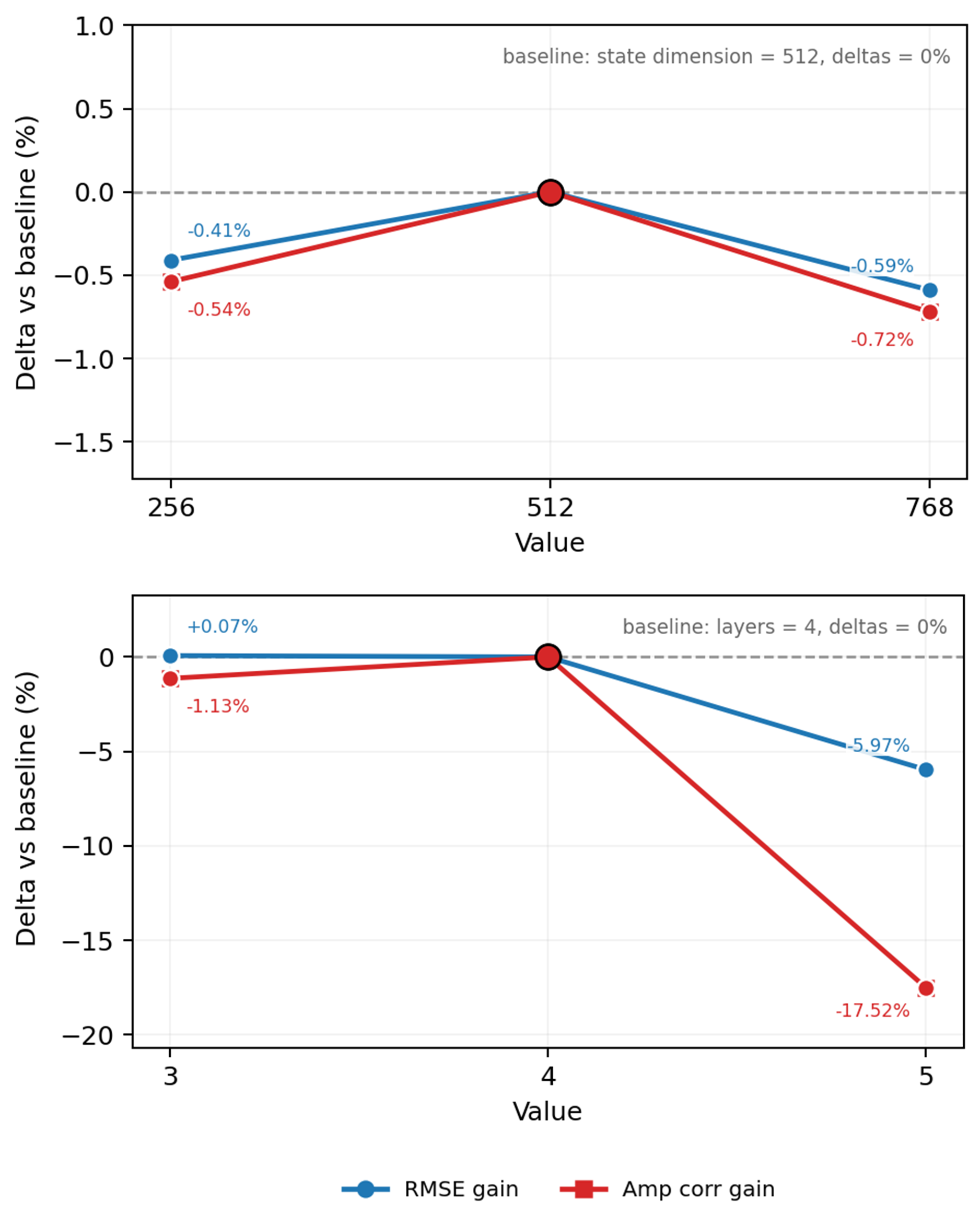}
  \caption{
    Capacity ablation for the high-capacity baseline (Stage-0).
    The top panel varies state dimension (baseline: 512); the bottom
    panel varies layer count (baseline: 4).
    Values show percentage deltas relative to the baseline for RMSE
    and amplitude-correlation gains; negative means worse.
    Alternative configurations were sub-optimal or comparable,
    supporting the chosen baseline.
  }
  \label{fig:baseline-ablation}
\end{figure}
The staged design separates architecture, regularization, and
loss-weight effects. Stage~I keeps the loss fixed and sweeps state
dimension \(d_{\mathrm{state}}\in\{256,512,768\}\),
layers \(L\in\{3,4,5\}\), dropout \(\{0,0.05,0.10\}\),
and selectivity and skip on/off.
As illustrated in \Cref{fig:baseline-ablation}, increasing capacity
beyond the baseline reference does not yield a consistent improvement.
The best Stage~I run instead changes regularization, selecting dropout
\(=0.10\) while retaining the baseline loss weights in
\Cref{tab:teacher-loss-settings}.
This improves RMSE by \(+1.15\%\) and amplitude correlation by
\(+3.10\%\), indicating that regularization is a more productive
first-order adjustment than simply increasing state dimension or depth.
Stage~II begins the loss-design sweep over the learning rate \(\eta\),
\(w_{\log},w_{\mathrm{ac}},w_{\mathrm{edge}}\), dropout and selectivity.
The sweep perturbs the amplitude-structure terms in
\(\mathcal{L}_{\mathrm{AF}}\).
The selected run uses dropout \(=0.10\), enables selectivity and
increases the amplitude-correlation weight from the Stage-0 value
\(w_{\mathrm{ac}}=0.20\) to \(w_{\mathrm{ac}}=0.22\).
Among the stage representatives in \Cref{tab:teacher-performance},
this yields the strongest RMSE improvement,
\(\Delta_R=+1.18\%\), while also increasing amplitude correlation by
\(\Delta_C=+4.36\%\).
Stages~III and IV refine the loss surface locally.
Stage~III sweeps
\(\eta\in\{3.8,4.0,4.2\}\times10^{-4}\) and
\(w_{\log}\in\{0.20,0.22\}\).
The selected setting, \(\eta=3.8\times10^{-4}\) and
\(w_{\log}=0.20\), yields only a modest gain over the baseline.
%This suggests that log-amplitude calibration alone is insufficient
%for preserving focused azimuth structure.
%
Stage~IV sweeps
\(\eta,w_{\log},w_{\mathrm{ac}},w_{\mathrm{edge}}\) and selects
\(\eta=3.8\times10^{-4}\), \(w_{\log}=0.24\),
\(w_{\mathrm{ac}}=0.20\) and \(w_{\mathrm{edge}}=0.01\),
increasing amplitude correlation to \(\Delta_C=+5.28\%\).
%This supports the interpretation that azimuth focusing quality depends
%on balancing log-amplitude calibration, amplitude-correlation alignment,
%and edge-sensitive structural terms.
%
The final stage sweeps dropout, selectivity, \(\eta\), the weights
\(w_c,w_{\log},w_{\mathrm{ac}}\),
\(w_{\mathrm{tail}},w_{\nabla},w_{\mathrm{psd}},
w_{\mathrm{fw}},w_{\mathrm{edge}}\), and adaptive reweighting.
The search covers local-basin, selectivity-correlation,
learning-rate-transfer, secondary-loss, adaptive-reweighting,
high-edge, and amplitude-correlation/edge-balanced families.
The final loss settings are reported in
\Cref{tab:teacher-loss-settings}.
As shown in \Cref{tab:teacher-performance}, the selected teacher
achieves the highest amplitude correlation and complex coherence
among the stage representatives while still improving RMSE
relative to the Stage-0 baseline.

%% file: tables/teacher_student_comparison.tex
\begin{table*}[t]
\centering
\scriptsize

\caption{
Teacher--student comparison on the test set
(\(n=25{,}600\) complex samples) and final loss/distillation hyperparameters.
%Panel A reports scalar reconstruction error, amplitude calibration, complex
coherence, and phase fidelity.
%Panel B reports the ground-truth-aligned
%azimuth-focusing loss weights and the student-only distillation coefficients.
The student remains close to the teacher in pointwise error metrics, but loses
substantial correlation and coherence.%, indicating that the main distillation gap is
%structured complex fidelity rather than aggregate RMSE.
}
\label{tab:teacher_student_distillation}

\setlength{\tabcolsep}{3.0pt}
\renewcommand{\arraystretch}{1.0}

\textbf{A. Test-set reconstruction and coherence metrics}

\vspace{2pt}

\begin{tabular*}{\textwidth}{
@{\extracolsep{\fill}}
l r r r
@{\hspace{10pt}}
l r r r
@{}
}
\toprule
\textbf{Metric}
& \textbf{Teacher}
& \textbf{Student}
& \textbf{\(\Delta\)}
&
\textbf{Metric}
& \textbf{Teacher}
& \textbf{Student}
& \textbf{\(\Delta\)} \\
\midrule

RMSE (\(\downarrow\))
& \textbf{7753} & 7920 & \(+2.16\%\)
&
Complex coh. (\(\uparrow\))
& \textbf{0.303} & 0.147 & \(-51.5\%\) \\

Amp. corr. (\(\uparrow\))
& \textbf{0.222} & 0.099 & \(-55.3\%\)
&
Phase coh. (\(\uparrow\))
& \textbf{0.240} & 0.107 & \(-55.5\%\) \\

Mean amp. err. (\(\downarrow\))
& \textbf{0.080} & 0.148 & \(+85.3\%\)
&
Phase MAE (\(\downarrow\))
& \textbf{76.2} & 84.8 & \(+11.2\%\) \\

P95 amp. err. (\(\downarrow\))
& 0.069 & \textbf{0.017} & \(-75.7\%\)
&
Phase RMSE (\(\downarrow\))
& \textbf{91.5} & 99.3 & \(+8.51\%\) \\

\bottomrule
\end{tabular*}

\vspace{0.75em}

\textbf{B. Final teacher/student loss and distillation hyperparameters}

\vspace{2pt}

\setlength{\tabcolsep}{2.2pt}

\begin{tabular*}{\textwidth}{
@{\extracolsep{\fill}}
l r r r r r r r r r c c r r r r r
@{}
}
\toprule
\textbf{Model}
& \(\boldsymbol{w_c}\)
& \(\boldsymbol{w_{\log}}\)
& \(\boldsymbol{w_{\mathrm{ac}}}\)
& \(\boldsymbol{w_{\mathrm{tail}}}\)
& \(\boldsymbol{w_{\nabla}}\)
& \(\boldsymbol{w_{\mathrm{psd}}}\)
& \(\boldsymbol{w_{\mathrm{fw}}}\)
& \(\boldsymbol{w_{\mathrm{edge}}}\)
& \(\boldsymbol{\epsilon}\)
& \textbf{PSD band}
& \(\boldsymbol{W_{\mathrm{info}}}\)
& \(\boldsymbol{\lambda_{\mathrm{KD}}}\)
& \(\boldsymbol{\alpha_c}\)
& \(\boldsymbol{\alpha_{\ell}}\)
& \(\boldsymbol{\alpha_{\rho}}\)
& \(\boldsymbol{\alpha_{\phi}}\) \\
\midrule

Teacher
& 0.04
& 0.18
& 0.18
& 0.12
& 0.24
& 0.04
& 0.04
& 0.08
& \(10^{-3}\)
& \([0.15,0.50]\)
& \(0.75+0.50\,\widetilde{I}\)
& 0
& 0
& 0
& 0
& 0 \\

Student
& 0.35
& 0.08
& 0.08
& 0.03
& 0.02
& 0.01
& 0.00
& 0.005
& \(10^{-3}\)
& \([0.15,0.50]\)
& same as teacher
& 0.05
& 0.25
& 0
& 0.08
& 0.05 \\

\bottomrule
\end{tabular*}

\vspace{0.35em}

\begin{minipage}{0.98\textwidth}
\emph{Notes.}
For \(\downarrow\) metrics, positive \(\Delta\) indicates larger error; for
\(\uparrow\) metrics, negative \(\Delta\) indicates lower fidelity. The amplitude
errors are absolute deviations from the ideal amplitude ratio of one.
\(W_{\mathrm{info}}\) is the per-strip informativeness weight, denoted \(W\) in
the experiment logs (see Appendix A).
\end{minipage}

\end{table*}

%% file: sec/5_discussion.tex
\section{Discussion}
\label{sec:discussion}

The central result of this paper is that, after standard front-end preprocessing, a substantial portion of the azimuth-focusing operator can be represented as a compact causal dynamical system. We compare our proposed method against two implementations of the RDA algorithm, one being the commonly used ''batched" RDA where all operations are performed in batches, and the other is a version of the RDA which relies on buffering a window of SAR history that can perform focusing of a single range line at a time that we call the linewise RDA: the comparison of these results are presented in Table~\ref{tab:complexity-comparison-main}. We state all our assumptions and derivations in Appendix~\ref{app:complexity}.
Due to the fact our model uses fixed memory footprint SSM layers as its backbone and does not require buffering data in the azimuth direction, we obtain a far superior memory footprint to both baselines; our online formulation also enables drastically reduced data-in-to-data-out focusing latency. We show detailed computational complexity and FLOPs derivations in Appendices ~\ref{app:complexity} and ~\ref{app:flops}. 
We also present extended discussion including limitations, failure cases, operational considerations and robustness in Appendix ~\ref{app:discussion}. 
\begin{table}[t]
  \centering
  \caption{Complexity, latency, memory, and compute comparison.
  Lower latency, memory, and GFLOPs per row are better.}
  \label{tab:complexity-comparison-main}

  \setlength{\tabcolsep}{3pt}
  \renewcommand{\arraystretch}{1.15}

  \resizebox{\columnwidth}{!}{%
  \begin{tabular}{@{}llccccc@{}}
    \toprule
    \textbf{Method} &
    \textbf{Complexity} &
    \makecell{\textbf{Line}\\\textbf{(ms)}$^{*}$} &
    \makecell{\textbf{Scan}\\\textbf{(ms)}$^{*}$} &
    \textbf{Mem.} &
    \makecell{\textbf{GFLOPs}\\\textbf{/scan}} &
    \makecell{\textbf{GFLOPs}\\\textbf{/row}} \\
    \midrule

    Traditional RDA &
    $\mathcal{O}(N_aN_r\log(N_aN_r))$ &
    N/A$^{\dagger}$ &
    \textbf{27,444} &
    16 GB &
    \textbf{122} &
    N/A$^{\dagger}$ \\

    Linewise RDA &
    $\mathcal{O}(N_aN_rN_b\log(N_bN_r))$ &
    857 (\textit{+243})$^{\ddagger}$ &
    17,140,000 &
    780 MB &
    45,400 &
    2.27 \\

    \rowcolor{gray!12}
    \textit{This work} &
    $\mathbf{\mathcal{O}(N_aN_r\log N_r)}$ &
    \textbf{15.1} &
    302,000 &
    \textbf{6 MB} &
    324 &
    \textbf{0.016} \\

    \bottomrule
  \end{tabular}%
  }

  \vspace{3pt}

  \begin{minipage}{\columnwidth}
    \footnotesize
    $^{*}$Latency measured on a single core of an AMD EPYC 7343 CPU.\\
    $^{\dagger}$Cannot process individual range lines online.\\
    $^{\ddagger}$Processing starts only after the history buffer is filled,
    adding delay beyond pure compute time.
  \end{minipage}
\end{table}

\subsection{Limitations and failure cases}
Our study is intentionally narrow in operating regime. The current formulation assumes a linear synthetic aperture: stripmap acquisition, broadside geometry, modest angular extent, and front-end preprocessing that already makes the azimuth task well conditioned. The deployed model is therefore not yet a drop-in replacement for wide-squint, spotlight, or heavily motion-corrupted collections.
Failure cases are most likely when those assumptions are violated. Large residual motion errors, stronger-than-expected range migration, unseen PRF or bandwidth regimes, and scene statistics far from Maya4 could all cause defocus, amplitude bias, or temporal lag in the emitted rows. 
%In addition, because StreamAZ is causal, it cannot exploit arbitrarily large future context in the way a fully offline batch imager can.
%, so some loss of fidelity relative to the best offline methods should be expected in challenging scenes. 
We include a more in depth discussion of failure cases and robustness in appendices \ref{app:sensitivity-robustness}, \ref{app:limits-and-failure} and \ref{app:operational-considerations}.

%% file: sec/appendix.tex
\clearpage
\appendix

\section{Reproducibility Details and Loss Implementation}
\label{app:loss-implementation}

\paragraph{Compute environment.}
Training was run on PBS \texttt{gpu4\_std} CUDA nodes with one GPU, \(16\) CPU
cores, and \(32\)GB RAM per training job. Inference and metric evaluation used
one GPU, \(16\) CPU cores, and up to \(64\)GB RAM. These resources describe model
training and offline evaluation compute only.

\subsection{Ground-Truth-Aligned Azimuth-Focusing Loss}
\label{app:gt-loss}

All loss terms are evaluated on one-dimensional azimuth strips of length \(L\).
Let \(\hat z\in\mathbb{C}^{L}\) denote the model prediction and
\(z\in\mathbb{C}^{L}\) the target strip in the normalized training space. The
physical-scale strips are obtained by inverse normalization,
\[
  \hat u = T^{-1}(\hat z),
  \qquad
  u = T^{-1}(z),
\]
where \(T^{-1}\) is the inverse of the normalization applied during data loading.
We define physical amplitudes and log-amplitudes as

\begin{gather*}
  \hat a_j = |\hat u_j|, \\
  a_j = |u_j|, \\
  \hat \ell_j = \log(\hat a_j+\epsilon), \\
  \ell_j = \log(a_j+\epsilon)
\end{gather*}

The implemented ground-truth-aligned objective is

\begin{equation}
\begin{split}
\mathcal L_{\mathrm{GT}}
&= w_c\mathcal L_c
 + w_{\log}\mathcal L_{\log}
 + w_{\mathrm{ac}}\mathcal L_{\mathrm{ac}} \\
&\quad + w_{\mathrm{tail}}\mathcal L_{\mathrm{tail}}
 + w_{\mathrm{grad}}\mathcal L_{\mathrm{grad}}
 + w_{\mathrm{psd}}\mathcal L_{\mathrm{psd}} \\
&\quad + w_{\mathrm{fw}}\mathcal L_{\mathrm{fw}}
 + w_{\mathrm{edge}}\mathcal L_{\mathrm{edge}}.
\end{split}
\label{eq:app-gt-loss}
\end{equation}

The complex data-fidelity term is computed in the normalized complex field,

\begin{equation}
\begin{aligned}
\mathcal L_c
&= \frac{1}{L}\sum_{j=1}^{L}|\hat z_j-z_j| \\
&= \frac{1}{L}\sum_{j=1}^{L}
\sqrt{(\hat r_j-r_j)^2+(\hat i_j-i_j)^2}
\end{aligned}
\label{eq:app-complex-loss}
\end{equation}

where \(\hat z_j=\hat r_j+\mathrm{i}\hat i_j\) and
\(z_j=r_j+\mathrm{i}i_j\). The remaining terms operate on inverse-normalized
physical amplitudes or their log-amplitudes.

The log-amplitude term is
\begin{equation}
\mathcal L_{\log}
=
\frac{1}{L}\sum_{j=1}^{L}|\hat \ell_j-\ell_j|
\label{eq:app-log-loss}
\end{equation}
Amplitude-shape agreement is measured by the Pearson correlation of physical
amplitudes,

\begin{equation}
\begin{aligned}
\rho_a
&=
\frac{\sum_{j=1}^{L}(\hat a_j-\bar{\hat a})(a_j-\bar a)}
{\sqrt{
\sum_{j=1}^{L}(\hat a_j-\bar{\hat a})^2
\sum_{j=1}^{L}(a_j-\bar a)^2
}+\epsilon},
\\
\mathcal L_{\mathrm{ac}}
&=
1-\operatorname{clip}(\rho_a,-1,1)
\end{aligned}
\label{eq:app-amplitude-correlation}
\end{equation}

The tail term penalizes mismatch in rare high-amplitude responses through upper
amplitude quantiles,
\begin{equation}
\mathcal L_{\mathrm{tail}}
=
\left|
\log
\frac{q_{0.95}(\hat a)+\epsilon}{q_{0.95}(a)+\epsilon}
\right|
+
\frac{1}{2}
\left|
\log
\frac{q_{0.99}(\hat a)+\epsilon}{q_{0.99}(a)+\epsilon}
\right|.
\label{eq:app-tail-loss}
\end{equation}
The first-order azimuth-gradient term matches local log-amplitude changes,
\begin{equation}
\mathcal L_{\mathrm{grad}}
=
\frac{1}{L-1}
\sum_{j=1}^{L-1}
\left|
(\hat \ell_{j+1}-\hat \ell_j)
-
(\ell_{j+1}-\ell_j)
\right|.
\label{eq:app-gradient-loss}
\end{equation}

\paragraph{PSD-band term.}
Let \(f_k=\mathrm{rfftfreq}(L)_k\) denote the normalized real-FFT frequencies.
The PSD loss is evaluated only on the band
\begin{equation}
\mathcal B_{\mathrm{PSD}}
=
\{f_k:f_{\min}\leq f_k\leq f_{\max}\}.
\label{eq:app-psd-band}
\end{equation}
For a log-amplitude strip \(\ell\), define

\begin{equation}
\begin{aligned}
P_k(\ell)
&=
\left|
\mathrm{rFFT}(\ell-\bar\ell)_k
\right|^2,
\\
\tilde P_k(\ell)
&=
\frac{P_k(\ell)}
{\sum_{m\in\mathcal B_{\mathrm{PSD}}}P_m(\ell)+\epsilon}.
\end{aligned}
\label{eq:app-normalized-psd}
\end{equation}

The PSD-band discrepancy is
\begin{equation}
\mathcal L_{\mathrm{psd}}
=
\frac{1}{|\mathcal B_{\mathrm{PSD}}|}
\sum_{k\in\mathcal B_{\mathrm{PSD}}}
\left|
\tilde P_k(\hat\ell)-\tilde P_k(\ell)
\right|.
\label{eq:app-psd-loss}
\end{equation}
In the experiments, \([f_{\min},f_{\max}]=[0.15,0.50]\) cycles/sample.

\paragraph{Focus-width term.}
The focus-width proxy is computed from the normalized autocorrelation of the
centered log-amplitude strip. With zero-padding length \(2L\),
\begin{equation}
A_{\ell}(\tau)
=
\frac{
\mathcal F^{-1}
\left(
|\mathcal F(\ell-\bar\ell;2L)|^2
\right)_{\tau}
}{
\mathcal F^{-1}
\left(
|\mathcal F(\ell-\bar\ell;2L)|^2
\right)_{0}
+\epsilon
}.
\label{eq:app-autocorrelation}
\end{equation}
The focus width is the first non-negative lag at which the normalized
autocorrelation falls below \(0.5\),

\begin{equation}
\begin{aligned}
FW(\ell)
&=
\min\{\tau:A_{\ell}(\tau)<0.5\},
\\
\mathcal L_{\mathrm{fw}}
&=
\frac{|FW(\hat\ell)-FW(\ell)|}{FW(\ell)+\epsilon}.
\end{aligned}
\label{eq:app-focus-width-loss}
\end{equation}

This term acts as a compact differentiable proxy for mainlobe broadening in the
log-amplitude response.

\paragraph{Second-order edge term.}
The edge term matches the second finite difference of the log-amplitude strip,

\begin{equation}
\begin{aligned}
\mathcal L_{\mathrm{edge}}
&= \frac{1}{L-2}
\sum_{j=1}^{L-2}
\Biggl|
(\hat\ell_{j+2}-2\hat\ell_{j+1}+\hat\ell_j) \\
&\quad -
(\ell_{j+2}-2\ell_{j+1}+\ell_j)
\Biggr|.
\end{aligned}
\label{eq:app-edge-loss}
\end{equation}

It penalizes mismatched local curvature and suppresses both excessive smoothing
and oscillatory ringing.

\subsection{Strip-Informativeness Reweighting}
\label{app:informativeness-weight}

To avoid treating low-structure and high-structure strips identically, each strip
in a mini-batch is assigned an informativeness weight. For strip \(i\), define

\begin{equation}
\begin{aligned}
G_i
&=
\frac{1}{L-1}
\sum_{j=1}^{L-1}
|\ell_{i,j+1}-\ell_{i,j}|,
\\
H_i
&=
\frac{1}{|\mathcal B_{\mathrm{PSD}}|}
\sum_{k\in\mathcal B_{\mathrm{PSD}}}
P_k(\ell_i),
\end{aligned}
\label{eq:app-info-components}
\end{equation}

and

\begin{equation}
\begin{aligned}
I_i
&=
G_i+\frac{1}{2}H_i,
\\
\eta_i
&=
\frac{I_i-\min_m I_m}{\max_m I_m-\min_m I_m+\epsilon}.
\end{aligned}
\label{eq:app-info-normalization}
\end{equation}

The final per-strip weight is
\begin{equation}
W_i
=
0.75+0.50\,\eta_i,
\qquad
W_i\in[0.75,1.25].
\label{eq:app-info-weight}
\end{equation}
The batch loss is the mean of the weighted strip losses,
\begin{equation}
\mathcal L_{\mathrm{GT}}^{\mathrm{batch}}
=
\frac{1}{B}
\sum_{i=1}^{B}
W_i\,
\mathcal L_{\mathrm{GT}}^{(i)}.
\label{eq:app-weighted-batch-loss}
\end{equation}
The PSD band used for both \(H_i\) and \(\mathcal L_{\mathrm{psd}}\) is computed
from normalized frequencies returned by \texttt{torch.fft.rfftfreq}, retaining
\(0.15\leq f\leq0.50\).

\subsection{Teacher--Student Distillation Loss}
\label{app:distillation-loss}

Let \(\hat z_s\) be the student prediction, \(\hat z_t\) the frozen teacher
prediction, and \(z\) the ground-truth target. The student objective combines the
ground-truth-aligned focusing loss with a teacher-output distillation penalty,
\begin{equation}
\mathcal L_{\mathrm{student}}
=
\mathcal L_{\mathrm{GT}}(\hat z_s,z)
+
\lambda_{\mathrm{KD}}\,
\mathcal L_{\mathrm{KD}}(\hat z_s,\hat z_t).
\label{eq:app-student-loss}
\end{equation}

The implemented distillation block is

\begin{equation}
\begin{aligned}
\mathcal L_{\mathrm{KD}}
&= \alpha_c\mathcal L_{c,\mathrm{KD}}(\hat z_s,\hat z_t)
 + \alpha_{\ell}\mathcal L_{\log}(\hat z_s,\hat z_t) \\
&\quad + \alpha_{\rho}\mathcal L_{\mathrm{ac}}(\hat z_s,\hat z_t)
 + \alpha_{\phi}\mathcal L_{\phi}(\hat z_s,\hat z_t).
\end{aligned}
\label{eq:app-kd-loss}
\end{equation}

Here \(\mathcal L_{\log}\) and \(\mathcal L_{\mathrm{ac}}\) are evaluated by
treating the teacher output as the target in the definitions above. The direct
complex-field distillation anchor is implemented as an MSE term,
\begin{equation}
\mathcal L_{c,\mathrm{KD}}(\hat z_s,\hat z_t)
=
\frac{1}{L}
\sum_{j=1}^{L}
|\hat z_{s,j}-\hat z_{t,j}|^2.
\label{eq:app-kd-complex-mse}
\end{equation}
This distinguishes the KD anchor from the ground-truth complex term in
Eq.~\eqref{eq:app-complex-loss}, which uses the complex magnitude error.

The phase term compares unit complex vectors and weights phase errors by the
teacher amplitude:

\begin{equation}
\begin{aligned}
\mathcal L_{\mathrm{KD}}
&= \alpha_c\mathcal L_{c,\mathrm{KD}}(\hat z_s,\hat z_t)
 + \alpha_{\ell}\mathcal L_{\log}(\hat z_s,\hat z_t) \\
&\quad + \alpha_{\rho}\mathcal L_{\mathrm{ac}}(\hat z_s,\hat z_t)
 + \alpha_{\phi}\mathcal L_{\phi}(\hat z_s,\hat z_t).
\end{aligned}
\label{eq:app-kd-loss}
\end{equation}

\begin{equation}
\begin{aligned}
m_j
&=
\frac{
|\hat z_{t,j}|^{p_{\phi}}
}{
\frac{1}{L}\sum_{m=1}^{L}|\hat z_{t,m}|^{p_{\phi}}+\epsilon_{\phi}
},
\\[6pt]
\mathcal L_{\phi}
&=
\frac{
\sum_{j=1}^{L}m_j|v_{s,j}-v_{t,j}|^2
}{
\sum_{j=1}^{L}m_j+\epsilon_{\phi}
}
\end{aligned}
\label{eq:app-phase-kd-loss}
\end{equation}

This formulation emphasizes teacher phase where the teacher has non-negligible
amplitude and avoids over-weighting phase in near-zero-amplitude regions.

\subsection{Training Details}
\label{app:training_details}
\paragraph{Training--deployment transition.}
streamAZ uses the same learned S4D parameters in two evaluation modes. During training, the
student is evaluated in convolutional mode on finite azimuth strips. The S4D convolution
kernel is computed with FFT-based zero-padded linear convolution using padding length
$2L$, and the output is cropped back to the original strip length $L$. This avoids circular
wrap-around while preserving the causal zero-state boundary condition used by the
recurrent processor.

The recurrent representation of S4D layers allows the model to be run in a way such that the output of the current time-step relies only on the current time step and hidden state, allowing for a memory footprint which is independent of the aperture length. We represent each pulse as two real channels, representing real and imaginary parts, and normalize inputs to improve numerical stability.
At each step the model produces a complex estimate for the currently resolvable image line. 

\subsection{Training and Evaluation Protocol}
\label{subsec:training-evaluation}

Student training uses single-range-bin azimuth strips of length \(L=1000\), sampled
in row order with \texttt{patch\_size=[1000,1]} and
\texttt{stride=[1000,1]}. Each mini-batch contains \(61\) strips, with gradients
accumulated over two mini-batches for an effective batch size of \(122\). We do
not use truncated backpropagation or sequence chunking: each strip is processed as
one complete \(1000\)-step sequence. Inputs and targets are normalized with an
origin-preserving complex scaling: for each component \(c\in\{\Re,\Im\}\), both
range-compressed inputs and azimuth-compressed targets use
\(\tilde c=c/2000\), with inverse transform \(c=2000\tilde c\). The symmetric
bounds \([-2000,2000]\) preserve \(0+0\mathrm{i}\) exactly. Predictions are
inverse-normalized before visualization and before computing all reported metrics.
The student is optimized with Adam using betas \((0.9,0.999)\), initial learning
rate \(4\times10^{-4}\), three warm-up epochs, cosine decay, gradient clipping at
\(0.5\), a maximum budget of \(300\) epochs, and early stopping with patience
\(40\) and \(\texttt{min\_delta}=10^{-5}\). The reported student checkpoint is
from epoch \(57\), with saved optimizer learning rate \(10^{-5}\). Data splitting
and patch sampling use seed \(42\); teacher training additionally fixes the
Python, NumPy, PyTorch, and CUDA seeds to \(42\).

\subsection{Model architecture details}

\begin{table}[h]
\caption{Model architecture hyperparameters for the teacher and student networks.}
\label{tab:architecture}
\centering
\begin{tabular}{lcc}
\toprule
 & \textbf{Teacher} & \textbf{Student} \\
\midrule
Number of S4D layers       &  4 & 4 \\
Hidden State Size          &  512 & 8 \\
Input Dimension            & 3  & 3 \\
Position embedding dim.    & Range & Range \\
Initialization             & S4D Lin & S4D Lin \\
Learning Rate              & 0.0005 & 0.002 \\
\bottomrule
\end{tabular}
\end{table}

\section{Downstream Tasks Details}

\subsection{CFAR Vessel Detection}

\label{sec:cfar}                                    
  To localize vessel candidates in the focused SAR intensity image we use a                                            
  two-dimensional cell-averaging constant false alarm rate (CA-CFAR) detector,                                         
  implemented as a streaming filter that consumes one range line at a time.                                            
  This matches the line-sequential nature of our front-end and back-end                                                
  processing pipeline and bounds the detector's working set to a small number                                          
  of azimuth lines rather than the full scene.                                                                         
                                                                                                                       
  \paragraph{Window geometry.}                                                                                         
  For each cell-under-test (CUT) at position $(i, j)$ in the intensity image
  $x_{i,j} = |s_{i,j}|^{2}$, we define two centered square windows: an outer                                           
  window of side $2(G + T) + 1$ and an inner (guard) window of side $2G + 1$,                                          
  where $G$ is the guard half-width and $T$ is the training half-width. The                                            
  training set $\mathcal{T}_{i,j}$ consists of the cells in the outer window                                           
  that are not in the inner window, so that                                                                            
  $|\mathcal{T}| = N_\text{train} = (2(G+T)+1)^{2} - (2G+1)^{2}$.
  We use $G = 10$ and $T = 28$, giving $N_\text{train} = 5{,}776$ training                                             
  cells.                                                                                                               
                                                                                                                       
  \paragraph{Detection rule.}                                                                                          
  Under the standard assumption that homogeneous clutter intensity follows an
  exponential distribution, the maximum-likelihood noise estimate is the                                               
  training-cell mean                                                                                                   
  $\hat{\mu}_{i,j} = \frac{1}{N_\text{train}} \sum_{(p,q) \in \mathcal{T}_{i,j}} x_{p,q}$,                             
  and the CA-CFAR decision compares the CUT to a scaled version of this                                                
  estimate,                                                                                                            
\begin{equation}
\begin{gathered}
x_{i,j} \;\gtrless\; \alpha \,\hat{\mu}_{i,j},
\\[6pt]
\alpha \;=\; N_\text{train}\!\left( P_\text{fa}^{-1/N_\text{train}} - 1 \right)
\end{gathered}
\end{equation}                                                                                           
  where the multiplier $\alpha$ is chosen so that the per-cell probability of                                          
  false alarm is $P_\text{fa}$ in the exponential-clutter model. We target                                             
  $P_\text{fa} = 10^{-6}$.                                                                              
  \paragraph{Streaming computation.}                                                
  A naive implementation evaluates a 2D sum over $\mathcal{T}_{i,j}$ at every
  pixel. We instead exploit the separability of the box sums. For each range                                           
  bin $j$, we maintain two running azimuth-column sums,                                                                
  $C^\text{outer}_{j}$ over the most recent $2(G+T)+1$ lines and                                                       
  $C^\text{inner}_{j}$ over the central $2G+1$ lines centered on the CUT                                               
  line. Once a CUT line has $G + T$ lines of azimuth context above                                              
  and below it, the 2D outer and inner box sums for every range bin on that                                            
  line are obtained in a single 1D pass along range by applying a length                                               
  $2(G+T)+1$ and a length $2G+1$ box filter, respectively, to the column-sum                                           
  vectors. The training sum is then the difference of these two box sums,                                              
  yielding the per-pixel noise estimate $\hat{\mu}_{i,j}$ and threshold for                                            
  the entire CUT line at constant cost per pixel and constant memory in                                                
  azimuth. The detector emits decisions with a fixed latency of $G + T = 38$                                           
  azimuth lines relative to the input stream.                                                              
  \paragraph{Boundary handling.}                                                                                       
  At the leading and trailing edges of the swath the outer window extends                                              
  beyond the available data. We use reflect padding in azimuth, mirroring                                              
  buffered lines about the first and last real lines, which is consistent                                              
  with the boundary mode of standard uniform-filter implementations and                                                
  avoids the bias that zero-padding would introduce into the noise estimate.                                           
                                              
  \paragraph{Post-processing.}                                                                                 
  The binary detection map is post-filtered by removing connected components                                           
  smaller than $K_\text{min} = 3$ pixels to suppress isolated single-pixel                                             
  detections that are dominated by speckle rather than persistent scatterers.                                          
  The resulting mask is the set of vessel candidates passed to downstream                                              
  evaluation
\subsection{Rapid Flood Detection}
\label{sec:water-segmentation}                                      
  To obtain a binary water mask from a focused single-look complex SAR scene, we exploit the fact that calm water surfaces behave as
  near-specular reflectors at C-band: incident energy is reflected away                                                
  from the sensor in the forward direction, so water pixels return very                                                
  little backscattered power and appear as dark regions in intensity                                                   
  imagery. Land surfaces, by contrast, scatter diffusely and yield  
  substantially higher returns. A simple intensity threshold therefore suffices to separate the two classes.                   
  \paragraph{Intensity formation.}                                  
  Given the complex-valued focused image $z \in \mathbb{C}^{H\times W}$,
  we form the per-pixel radar intensity                            
  \begin{equation}                                     
      I(x,y) \;=\; |z(x,y)|^2,                                
  \end{equation}                                                                                           
  which is proportional to the radar cross-section of the resolution
  cell (up to the system calibration constant, which we do not require).         
  \paragraph{Smoothing.}                       
In order to lower the impact of noise on the output of the segmentation map we convolve the intensity image with a $K\times K$ spatial kernel,    
  \begin{equation}                      
      \tilde{I}(x,y) \;=\; (h_K \ast I)(x,y),  
  \end{equation}                                                  
  where $h_K$ is either a uniform (boxcar) filter of side length K.        
  
  \paragraph{Thresholding.}                                                                                            
  A pixel is labelled as water when its normalised intensity falls below
  a fixed cutoff $\tau$,                                                                                               
  \begin{equation}                                                                                                     
      M(x,y) \;=\;                                                                                                     
      \mathbf{1}\!\left[\, I_{\mathrm{dB}}(x,y) < \tau \,\right],                                                      
  \end{equation}                                                                                                       
  with $\tau = -9\,\mathrm{dB}$ chosen empirically.                                                       
  \paragraph{Connected-component cleanup.}                         Scans produced with streamAZ or traditional methods tend to have noise present. As such, to minimize the impact of random noise on the quality of our segmentation map, we label the binary mask under                                                                 
  $8$-connectivity and discard every connected component whose area is                                                 
  smaller than $A_{\min} = 600$ pixels, retaining only contiguous                                                      
  water bodies of physically plausible size. The final mask is                                                         
  \begin{equation}                                                                                                     
      M^{\star}(x,y) \;=\;                                                                                             
      \mathbf{1}\!\left[\, (x,y) \in \mathcal{C}_i,\;                                                                  
      |\mathcal{C}_i| \geq A_{\min} \,\right],                                                                         
  \end{equation}
  where $\{\mathcal{C}_i\}$ are the connected components of $M$. 
  The resulting binary map $M^{\star}$ is used as a weak label for 
  downstream evaluation; the pipeline is fully unsupervised and
  deterministic, requiring only the focused complex image as input.

  \paragraph{Limitations}
  The implementation of our code for this downstream tasks was not strictly a linewise implementation, however in principle this task could be implemented in a linewise manner. If the connected-component cleanup step is removed, all other steps could be implemented on a small rolling buffer of a few azimuth lines.

\section{Computational Complexity Derivations}
\label{app:complexity}

In this appendix we present a computational complexity derivation for two baseline range-doppler algorithm implementations, as well our Online Processor, as below:
\begin{itemize}
    \item Batched RDA
    \item Linewise RDA
    \item Online Processor
\end{itemize}

In the computational complexity derivations below we make the following assumptions: Let $N_a$ and $N_r$ denote the number of azimuth (slow-time) and range (fast-time) samples, respectively, so that the raw phase-history data occupies an $N_a \times N_r$ complex-valued matrix. In all three algorithms, the range compression, range-cell migration correction (RCMC), and azimuth compression kernels, where used, are precomputed before data acquisition begins and therefore we do not count them as contributing to the run-time complexity counts reported here. It should be noted, within this appendix we consider operations in terms of complex operations. All FFT computations in this appendix are assumed to be using the standard radix-2 Cooley-Tukey FFT.

% -------------------------------------------------------------------
\subsection{Batched Range-Doppler Algorithm}
% -------------------------------------------------------------------

The standard RDA processes the entire $N_a \times N_r$ phase-history array in one pass - we refer to this as the "batched" RDA as it focuses SAR data in batches. Its stages proceed as follows.

\paragraph{Step 1: Range Compression}
First, each of the $N_a$ azimuth lines is independently transformed to the range-frequency domain. Following this, each of the $N_r$ range lines is also independently transformed to the azimuth frequency domain, yielding the full two-dimensional frequency representation of the scene. Finally, the precomputed range matched filter $H_r$ is applied via element-wise complex multiplication across the $N_a \times N_r$ grid.

\paragraph{Step 2: Range-Cell Migration Correction}
In the two-dimensional frequency domain, each sample undergoes a range shift correction by means of an elementwise multiplication with an RCMC kernel, $H_{rcmc}$, requiring a single complex multiply-accumulate operation per cell.

\paragraph{Step 3: Azimuth Compression} 
Following RCMC, an inverse FFT is applied along the range direction to return each azimuth line to the range-time domain, producing the Range-Doppler representation. The precomputed azimuth matched filter $H_a$ is then applied via elementwise complex multiplication in the Range-Doppler domain. Finally, an inverse FFT along the azimuth direction recovers the focused complex image.

We find that the total computational complexity of this algorithm in terms of $N_a$ and $N_r$ is $N_a N_r\bigl[2\log_2(N_a N_r)+3\bigr]$, giving an order of $\mathcal{O}(N_a N_r log(N_a Nr))$. We present a breakdown of the computational cost of each stage in Table~\ref{tab:rda_cost}

\begin{table*}[h]
    \centering
    \caption{Per-stage operation count for the Range-Doppler Algorithm with precomputed filters.}
    \label{tab:rda_cost}
    \vspace{4pt}
    \begin{tabular}{@{}clll@{}}
        \toprule
        \textbf{Step} & \textbf{Stage} & \textbf{Operation} & \textbf{Cost} \\
        \midrule
        1 & Range FFT           & $N_a$ FFTs of length $N_r$    & $N_a N_r \log_2 N_r$ \\
        2 & Azimuth FFT         & $N_r$ FFTs of length $N_a$    & $N_a N_r \log_2 N_a$ \\
        3 & Range compression   & Element-wise $\odot\,H_r$     & $N_a N_r$ \\
        4 & RCMC                & Element-wise $\odot\,H_{RCMC}$     & $3 \cdot N_a N_r$ \\
        5 & Range IFFT          & $N_a$ IFFTs of length $N_r$   & $N_a N_r \log_2 N_r$ \\
        6 & Azimuth compression & Element-wise $\odot\,H_a$     & $N_a N_r$ \\
        7 & Azimuth IFFT        & $N_r$ IFFTs of length $N_a$   & $N_a N_r \log_2 N_a$ \\
        \midrule
          & \textbf{Total}      &                               & $N_a N_r\bigl[2\log_2(N_a N_r)+3\bigr]$ \\
          & \textbf{Order} & & $\mathcal{O}(N_a N_r log(N_a Nr))$\\
        \bottomrule
    \end{tabular}
\end{table*}

% -------------------------------------------------------------------
\subsection*{Line-wise Range-Doppler Algorithm}
% -------------------------------------------------------------------

The batched RDA requires all $N_a$ range lines to be resident in memory simultaneously, since the azimuth FFT must span the full synthetic aperture (on the order of tens of thousands of pulses for Sentinel-1 stripmap data). A line-wise variant is possible, but not commonly used, and can be implemented if a certain amount of phase history is buffered; while this is impractical in most real systems, we include it here to provide a direct algorithmic baseline against which to compare our proposed method.

In such a method we maintain a rolling buffer of $N_b$ range lines simultaneously. The reason for doing this is that after the RC and RCMC step, depending on the squint angle, SAR information in the sample under processing is spread across many azimuth time steps. As such, to focus properly any given azimuth cell, it is required to have knowledge of the data corresponding to the time steps before and after this cell's occurrence. This value is related to the length of the synthetic aperture, $L_{sa}$:
\begin{equation}
    L_{sa} = R_o\cdot \Theta
\end{equation}
where $\Theta$ is the angular span of the SAR antenna and $R_o$ is the range of closest approach. Following this, we can approximate $\Theta$ as the ratio of the transmitted wavelength $\mu$ and the physical antenna length:
\begin{equation}
L_{ant}: \Theta \approx \frac{\mu}{L_{ant}}
\end{equation}
Finally, given that: 
\begin{equation}
N_{az} = \frac{L_{SA}}{\Delta x_{az}}
\end{equation}
where $\Delta x_{az}$ is the ground-projected size, we can obtain: 
\begin{equation}
N_{az} \approx \frac{R_o \mu}{L_{ant}\Delta x_{az}}
\end{equation}
resulting in an estimate of approximately 486 azimuth cells using Sentinel-1 characteristics. As such, we use a buffer of 486 cells looking forward and backward, resulting in a buffer of $N_b$ = 972. We assume this number for our computational complexity assumptions, however it should be noted that $R_0$ changes for different stripmap modes, and we present these values below in Table~\ref{tab:sentinel1_r0} for the reader's reference.

\begin{table*}[h]
    \centering
    \caption{Near-range slant range $R_0$ for Sentinel-1 stripmap beams at minimum orbit altitude ($\sim$698~km). These values contextualize the buffer-size estimate used in the linewise-RDA latency and complexity calculations.}
    \label{tab:sentinel1_r0}
    \vspace{4pt}
    \begin{tabular}{@{}cccc@{}}
        \toprule
        \textbf{Beam} & \textbf{Off-nadir angle [$^\circ$]} & \textbf{Incidence angle [$^\circ$]} & \textbf{$R_0$ [km]} \\
        \midrule
        S1 & 17.93 & 19.99 & 737.9 \\
        S2 & 21.00 & 23.45 & 753.8 \\
        S3 & 26.18 & 29.33 & 788.4 \\
        S4 & 30.87 & 34.71 & 829.8 \\
        S5 & 35.07 & 39.62 & 877.2 \\
        S6 & 37.53 & 42.53 & 910.6 \\
        \bottomrule
    \end{tabular}
\end{table*}

It should be noted that the total delay for such an implementation is equal to the processing latency summed with a buffering latency. The buffering latency is associated with the time it takes for the satellite to fill the buffer of phase history in the forward direction after the data for the range line of interest has been recorded but before the data associated with the range line of interest can start to be processed. For a SAR system with a sampling frequency of 2000Hz, this delay is around 0.246 seconds, as such we include this in our latency calculations.

\paragraph{Per-output-line computation cost.}
This formulation would use a sliding buffer of the most recent $N_b$ range-compressed pulses. Each time a new slow-time sample arrives, the following operations are performed:

\paragraph{Step 1: Range Compression}
Each of the $N_b$ range lines in the buffer are independently transformed to the range-frequency domain. However, this action only has to occur once per range-line, and after performed the results for this can be cached, such that for each timestep only the most recent range line needs to be converted to the range-frequency domain. Following this, each of the $N_r$ range bins are independently transformed to the azimuth-frequency domain, yielding a two-dimensional frequency representation $N_b$ slow-time timesteps of phase history. Next, the precomputed range matched filter $H_r$ is applied via element-wise complex multiplication across the $N_b \times N_r$ grid.

\paragraph{Step 2: Range-Cell Migration Correction} In the two-dimensional frequency domain, each sample undergoes a range shift that is corrected via elementwise multiplication with $H_{rcmc}$

\paragraph{Step 3: Azimuth Compression} Following RCMC, an inverse FFT is applied along the range direction to return each azimuth line to the range-time domain, producing the Range-Doppler representation. A final inverse FFT along azimuth recovers a partially focused complex image

\paragraph{Step 4: Extract Valid Data} Only one range row of data is valid and can be considered fully focused from the buffer output. Thus, we extract one valid range row of focused SAR data from the center of the output image and disregard the remaining data. Then, the previous steps are repeated every PRI, assuming $N_a$ pulses across the entire sample, this results in a total complexity of $N_r\bigl[\log_2 N_r +  N_b\bigl[3 + 2\log_2 N_r + \log_2 N_b \bigr]\bigr]$. A detailed breakdown of the computational complexity of each step is presented in Table~\ref{tab:lrda_cost}

\begin{table*}[h]
    \centering
    \caption{Per-stage operation count for the Line-wise Range-Doppler Algorithm. Each block of operations is executed once per incoming pulse and repeated $N_a$ times. $N_b$ is the rolling buffer size (constant for a given SAR mode).}
    \label{tab:lrda_cost}
    \footnotesize
    \begin{tabular}{@{}clll@{}}
        \toprule
        \textbf{Step} & \textbf{Stage} & \textbf{Operation} & \textbf{Cost (per pulse)} \\
        \midrule
        1 & Range FFT           & single FFT of length $N_r$    & $N_r \log_2 N_r$ \\
        2 & Azimuth FFT         & $N_r$ FFTs of length $N_b$    & $N_b N_r \log_2 N_b$ \\
        3 & Range compression   & Element-wise $\odot\,H_r$     & $N_b N_r$ \\
        4 & RCMC                & Element-wise $\odot\,H_{RCMC}$& $N_b N_r$ \\
        5 & Range IFFT          & $N_b$ IFFTs of length $N_r$   & $N_b N_r \log_2 N_r$ \\
        6 & Azimuth compression & Element-wise $\odot\,H_a$     & $N_b N_r$ \\
        7 & Azimuth IFFT        & $N_r$ IFFTs of length $N_b$   & $N_b N_r \log_2 N_b$ \\
        \midrule
          & \textbf{Per-pulse total} &                           & $N_r\bigl[\log_2 N_r +  N_b\bigl[3 + 2\log_2 N_r + \log_2 N_b \bigr]\bigr]$ \\
          & \textbf{Full-scene total} & $\times\; N_a$ pulses   & $N_a N_r\bigl[\log_2 N_r +  N_b\bigl[3 + 2\log_2 N_r + \log_2 N_b \bigr]\bigr]$ \\
          & \textbf{Order} & & \textbf{$\mathcal{O}(N_a N_r N_b \log(N_b N_r))$} \\
        \bottomrule
    \end{tabular}
\end{table*}

% -------------------------------------------------------------------
\subsection*{Line-wise State-Space Model}
% -------------------------------------------------------------------

Our proposed Online Processor uses an SSM operating in the recurrent mode: it maintains a hidden state per range cell that persists in dimension across slow-time steps. Each new pulse is range-compressed and the resulting $N_r$ range-cell values are fed into SSM instances with independent hidden states, which implicitly learn azimuth compression and RCMC without any explicit FFT along the azimuth direction. The SSM has fixed architecture parameters: $K$ layers, hidden dimension $H$, and channel number $N$. In our model we also use fully-connected layers, at the front and end of the model and within the S4D layers. The 1st fully connected layer requires 8 multiply accumulates and the other layers all require 4 multiply accumulates. As such, we will represent this fixed number of MACs per model inference as $X$. Given that the architecture of the model is fixed, our online processor only increases in algorithmic complexity in terms of the number of range bins $N_r$ and the number of slow-time steps $N_a$. However, we note that at smaller values of $N_r$ and $N_a$ that $X$ will be significant.

\paragraph{Step 1: Range Compression}
Perform one FFT of length $N_r$ on input range line. Then apply the precomputed matched filter by means of an elementwise multiply. Finally, return to range-time domain by performing IFFT of length $N_r$ on the range line

\paragraph{Step 2: SSM inference} We perform one neural network forward pass of our tiny model per range cell. We denote the inference cost of the tiny neural network as $X$ MACs - constant with respect to both $N_a$ and $N_r$ - so performing across $N_r$ range cells each timesteps results in computational complexity of $X \times N_r$.

As such, our total computational complexity comes out to $N_a N_r\bigl[1 + X + 2\log_2 N_r\bigr]$.

\begin{table*}[h]
    \centering
    \caption{Per-stage operation count for the Online Processor. Each block of operations is executed once per incoming pulse and repeated $N_a$ times.}
    \label{tab:osp_cost}
    \footnotesize
    \begin{tabular}{@{}clll@{}}
        \toprule
        \textbf{Step} & \textbf{Stage} & \textbf{Operation} & \textbf{Cost (per pulse)} \\
        \midrule
        1 & Range FFT           & single FFT of length $N_r$    & $N_r \log_2 N_r$ \\
        2 & Range compression   & Element-wise $\odot\,H_r$     & $N_r$ \\
        3 & Range IFFT          & $N_b$ IFFTs of length $N_r$   & $N_r \log_2 N_r$ \\
        4 & Model Inference & $N_r$ model inferences     & $X N_r$ \\

        \midrule
          & \textbf{Per-pulse total} &                           & $N_r\bigl[1 + X + 2\log_2 N_r\bigr]$ \\
          & \textbf{Full-scene total} & $\times\; N_a$ pulses   & $N_a N_r\bigl[1 + X + 2\log_2 N_r\bigr]$ \\
          & \textbf{Order} & & \textbf{$\mathcal{O}(N_a N_r\log(N_r))$} \\
        \bottomrule
    \end{tabular}
\end{table*}

It should be noted that for the version of our algorithm used in the experiments presented in this paper,  the operations of X is larger than $\log_2 N_r$, however since X is of a fixed size, as the data size of the SAR scan increases (portending the increase of $N_r$ and $N_a$), this scales as $\mathcal{O}(N_a N_r \log_2 N_r)$, which is particularly pertinent given the trend of SAR systems to larger and larger data volumes. In Table~\ref{tab:cost_summary} we summarize the results of our computational complexity analysis.

\begin{table*}[h]
    \centering
    \caption{Summary of total computational cost across methods.}
    \label{tab:cost_summary}
    \footnotesize
    \begin{tabular}{@{}lll@{}}
        \toprule
        \textbf{Method} & \textbf{Per-pulse total} & \textbf{Full-scene total}\\
        \midrule
        Batched RDA &
        -- &
        $N_a N_r\bigl[2\log_2(N_a N_r)+3\bigr]$ \\[6pt]
        Line-wise RDA &
        $N_r\bigl[\log_2 N_r + N_b\bigl[3 + 2\log_2 N_r + \log_2 N_b\bigr]\bigr]$ &
        $N_a N_r\bigl[\log_2 N_r + N_b\bigl[3 + 2\log_2 N_r + \log_2 N_b\bigr]\bigr]$ \\[6pt]
        Line-wise SSM &
        $N_r\bigl(2\log_2 N_r + 1 + X\bigr)$ &
        $N_a N_r\bigl(2\log_2 N_r + 1 + X\bigr)$  \\
        \bottomrule
    \end{tabular}
\end{table*}

\section{Computational Cost Analysis}
\label{app:flops}
 
This appendix derives the floating-point operation (FLOP) counts for our two SAR processing benchmarks and our proposed online processor.
For this comparison, we assume all three algorithms process a complex-valued SAR data matrix of size $N_a \times N_r = 20{,}000 \times 20{,}000$. This close to the size of a typical Sentinel-1 stripmap scan, although they tend to be long along the azimuth direction.
The linewise algorithms process one azimuth line at a time and are therefore applied $N_a = 20{,}000$ times to fully process the entire scan.
 
\subsection{Counting Conventions}
 
Throughout this appendix, each real-valued multiplication or addition is counted as one FLOP. By extension, a complex--complex multiplication $(a + bi)(c + di) = (ac - bd) + (ad + bc)i$ requires 4 real multiplies and 2 real adds $= 6$ FLOPs.
A complex--real multiplication $(a + bi) \cdot c = ac + bci$ requires 2 real multiplies $= 2$ FLOPs.
A complex addition requires 2 real adds $= 2$ FLOPs.
A radix-2 complex FFT of length $N$ requires approximately $5N\log_2 N$ real FLOPs (following \cite{cooley1965}).
All tensors use \texttt{complex64} (8 bytes per element: 2$\times$\texttt{float32}).
%--------------------------------------------------------------
\subsection{Batched RDA}
\label{sec:rda_batch}
 
A single iteration processes the full $20{,}000 \times 20{,}000$ complex matrix. The total steps are as follows:
 
\begin{enumerate}[nosep]
    \item Row-wise FFT, Column-wise FFT
    \item Three element-wise complex multiplies (RC, RCMC, AC filters)
    \item Row-wise IFFT, Column-wise IFFT
\end{enumerate}
 
\subsubsection{FLOPs}
 
\begin{table*}[h]
\centering
\caption{FLOP breakdown for batched RDA ($20{,}000 \times 20{,}000$).}
\label{tab:rda_batch_flops}
\footnotesize
\begin{tabular}{@{}lll@{}}
\toprule
\textbf{Operation} & \textbf{Expression} & \textbf{GFLOPs} \\
\midrule
FFT dim=1 & $20{,}000 \times 5 \times 20{,}000 \times \log_2(20{,}000)$ & 28.6 \\
FFT dim=0 & $20{,}000 \times 5 \times 20{,}000 \times \log_2(20{,}000)$ & 28.6 \\
Multiply (RC) & $20{,}000^2 \times 6$ & 2.4 \\
Multiply (RCMC) & $20{,}000^2 \times 6$ & 2.4 \\
Multiply (AC) & $20{,}000^2 \times 6$ & 2.4 \\
IFFT dim=1 & (same as FFT dim=1) & 28.6 \\
IFFT dim=0 & (same as FFT dim=0) & 28.6 \\
\midrule
\textbf{Total} & & \textbf{121.6} \\
\bottomrule
\end{tabular}
\end{table*}

%--------------------------------------------------------------
\subsection{Linewise RDA}
\label{sec:rda_line}
 
Here we also assume a buffer length $N_b$ of 972 as in Appendix~\ref{app:complexity}
Each iteration operates on a $972\times 20{,}000$ complex matrix and performs:
 
\begin{enumerate}[nosep]
    \item Row-wise FFT (dim=1, $N=20{,}000$, applied to 1 rows)
    \item Column-wise FFT (dim=0, $N=972$, applied to 20{,}000 columns)
    \item Element-wise complex multiply with range-compression filter
    \item Element-wise complex multiply with range-cell-migration-correction filter
    \item Element-wise complex multiply with azimuth-compression filter
    \item Row-wise IFFT (dim=1, $N=20{,}000$,
    applied to 972 rows)
    \item Column-wise IFFT (same cost as step 2)
\end{enumerate}
 
\subsubsection{FLOPs Per Iteration}
 
\begin{table*}[h]
\centering
\caption{FLOP breakdown per linewise RDA iteration ($972 \times 20{,}000$).}
\label{tab:rda_line_flops}
\begin{tabular}{@{}lll@{}}
\toprule
\textbf{Operation} & \textbf{Expression} & \textbf{GFLOPs} \\
\midrule
FFT dim=1 & $5 \times 20{,}000 \times \log_2(20{,}000)$ & 0.000143 \\
FFT dim=0 & $20{,}000 \times 5 \times 972 \times \log_2(972)$ & 0.965 \\
Multiply (RC filter) & $972 \times 20{,}000 \times 6$ & 0.0194 \\
Multiply (RCMC filter) & $972 \times 20{,}000 \times 6$ & 0.0194 \\
Multiply (AC filter) & $972 \times 20{,}000 \times 6$ & 0.0194 \\
IFFT dim=1 & $972 \times 20{,}000 \times \log_2(20{,}000)$  & 0.277 \\
IFFT dim=0 & (same as FFT dim=0) & 0.965 \\
\midrule
\textbf{Per-iteration total} & & \textbf{2.27} \\
\textbf{Per-scan total} & $:\times 20{,}000$ & \textbf{45400} \\
\bottomrule
\end{tabular}
\end{table*}

%--------------------------------------------------------------
\subsection{Online Processor}
\label{sec:op_line}
 
Each iteration operates on one range line ($1 \times 20{,}000$ complex) and performs:
\begin{enumerate}[nosep]
    \item Row-wise FFT, complex multiply with RC filter, row-wise IFFT (range compression)
    \item Tiny Neural network forward pass on all $N_r = 20{,}000$ range cells
\end{enumerate}

This process is then repeated in a linewise manner on each input range line of raw SAR data.

The Tiny Model consists of the following layers, applied sequentially:
 
\begin{center}
\texttt{fc1} $\rightarrow$ \texttt{ssm2} $\rightarrow$ act $\rightarrow$
\texttt{fc3} $\rightarrow$ \texttt{ssm4} $\rightarrow$ act $\rightarrow$
\texttt{fc5} $\rightarrow$ \texttt{ssm6} $\rightarrow$ act $\rightarrow$
\texttt{fc7} $\rightarrow$ \texttt{ssm8} $\rightarrow$ act $\rightarrow$
\texttt{fc9} $\rightarrow$ \texttt{fc10}
\end{center}

\begin{table*}[h]
\centering
\caption{Online Processor FLOPs estimate.}
\label{tab:ssm_dsp_flops}
\footnotesize
\begin{tabular}{@{}llll@{}}
\toprule
\textbf{Operation} & \textbf{Expression} & \textbf{FLOPs per Cell} & \textbf{MFLOPs per line}\\
\midrule
FFT dim=1 & $1 \times 5 \times 20{,}000 \times \log_2(20{,}000)$ & & 1.43 \\
Multiply (RC filter) & $20{,}000 \times 6$ & & 0.12 \\
IFFT dim=1 & (same as FFT) & & 1.43 \\
\midrule
\textbf{Front End Per-iteration total} & & & \textbf{2.98} \\
\midrule
\texttt{fc1} & FC input & 18 & \\
\texttt{ssm2} & S4D  & 146 & \\
act & LeakyReLU &  2 & \\
\texttt{fc3} & FC &  10 &\\
\texttt{ssm4} & S4D &  146 &\\
act & LeakyReLU & 2 & \\
\texttt{fc5} & FC & 10 & \\
\texttt{ssm6} & S4D & 146 & \\
act & LeakyReLU & 2 &  \\
\texttt{fc7} & FC & 10 & \\
\texttt{ssm8} & S4D &  146 & \\
act & LeakyReLU & 2 &  \\
\texttt{fc9} & FC & 10 & \\
\texttt{fc10} & FC & 10 & \\
\midrule
\textbf{Tiny Model Total} & & \textbf{660} & \textbf{13.2} \\
\midrule
\textbf{Online Processor Total} & & & \textbf{16.18} \\
\midrule \midrule
\textbf{Online Processor Per Scan Total} & & & \textbf{323600} \\
\bottomrule
\end{tabular}
\end{table*}

\section{Memory Footprint Calculations}

\subsection{Batched RDA}
 
Four tensors (\texttt{x}, \texttt{rc\_filter}, \texttt{rcmc\_filter}, \texttt{ac\_filter}), each $20{,}000 \times 20{,}000 \times 8 = 3.2$~GB, plus an FFT scratch buffer:
\begin{equation}
    M_{\text{RDA,batch}} \approx 5 \times 3.2 = \mathbf{16 \;\text{GB (peak)}}
\end{equation}

\subsubsection{Linewise RDA}
 
Each of the four tensors (\texttt{x}, \texttt{rc\_filter}, \texttt{ac\_filter}) occupies $972 \times 20{,}000 \times 8 = 156$~MB.
Including a scratch buffer for the in-place FFT:
\begin{equation}
    M_{\text{RDA,line}} \approx 5 \times 156 = \mathbf{780 \;\text{MB (peak)}}
\end{equation}

\subsubsection{Online Processor}
 
\begin{table*}[h]
\centering
\caption{Memory breakdown for linewise SSM.}
\label{tab:ssm_mem}
\footnotesize
\begin{tabular}{@{}llr@{}}
\toprule
\textbf{Component} & \textbf{Expression} & \textbf{Size} \\
\midrule
Signal tensors (\texttt{x}, \texttt{rc\_filter}) & $2 \times 20{,}000 \times 8$ & 320 KB \\
SSM state ($4$ layers $\times N_r \times [2,4]$ complex) & $4 \times 20{,}000 \times 2 \times 4 \times 8$ & 5.12 MB \\
Model parameters (real and complex) & $96\times8 + 48\times4$ & $<$1 KB \\
Embedding tensor ($N_r \times 4$, real) & $20{,}000 \times 4 \times 4$ & 320 KB \\
\midrule
\textbf{Total} & & $\mathbf{\approx 5.8}$ \textbf{MB} \\
\bottomrule
\end{tabular}
\end{table*}

\section{Dataset}
\begin{figure*}[h]
    \centering
    \includegraphics[width=\linewidth]{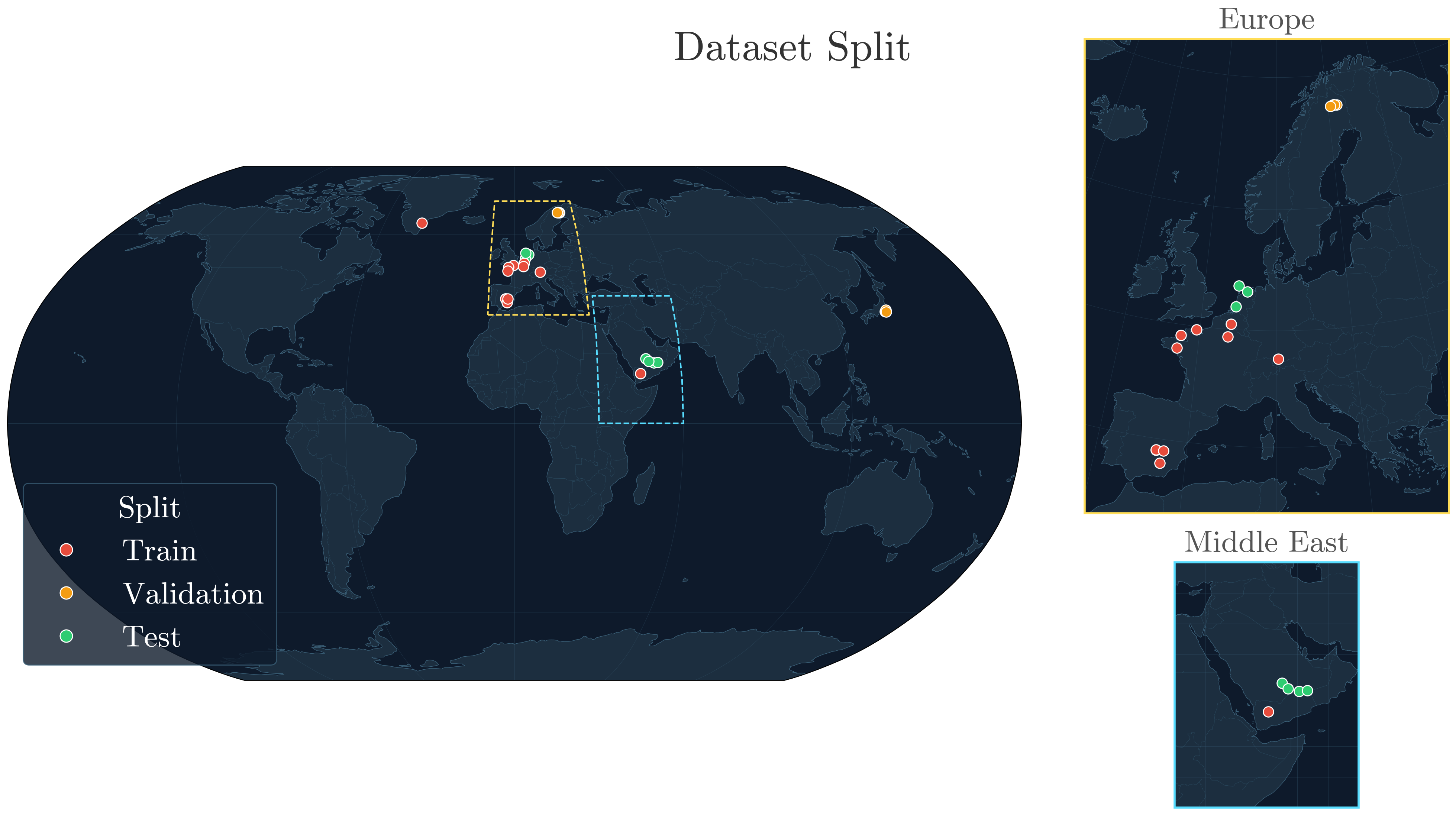}
    \caption{Maya4 subset scan locations map.}
    \label{fig:pt1_scan_locations}
\end{figure*}

The dataset on which our model was trained used a training-validation-testing split, with each of the splits being in geographically independent locations with both land and sea data to demonstrate the generalizability of our model. This dataset also included samples from all 6 stripmap modes used by the sentinel-1 satellites. We also present a map showing the geographical spread of our samples in figure ~\ref{fig:pt1_scan_locations}

\begin{table*}[h]
  \centering
  \caption{Dataset split into training, validation, and test scenes.}
  \label{tab:dataset-split}
  \footnotesize
  \setlength{\tabcolsep}{6pt}
  \begin{tabular}{cl}
    \toprule
    Split & Filename \\
    \midrule
    \multirow{16}{*}{Train}
      & \texttt{s1c-s1-raw-s-vv-20250417t025744-20250417t025817-001927-003c59.zarr} \\
      & \texttt{s1c-s1-raw-s-vv-20250417t025809-20250417t025828-001927-003c59.zarr} \\
      & \texttt{s1c-s2-raw-s-vv-20250403t062508-20250403t062540-001725-002fa3.zarr} \\
      & \texttt{s1c-s2-raw-s-vv-20250405t060531-20250405t060604-001754-00317b.zarr} \\
      & \texttt{s1c-s2-raw-s-vv-20250405t060556-20250405t060629-001754-00317b.zarr} \\
      & \texttt{s1c-s2-raw-s-vv-20250427t062507-20250427t062539-002075-004519.zarr} \\
      & \texttt{s1c-s4-raw-s-vv-20250329t061359-20250329t061431-001652-002b01.zarr} \\
      & \texttt{s1c-s4-raw-s-vv-20250329t061424-20250329t061459-001652-002b01.zarr} \\
      & \texttt{s1c-s4-raw-s-vv-20250417t091925-20250417t091957-001931-003c9c.zarr} \\
      & \texttt{s1c-s4-raw-s-vv-20250417t091950-20250417t092006-001931-003c9c.zarr} \\
      & \texttt{s1c-s5-raw-s-vv-20250416t052542-20250416t052609-001914-003b87.zarr} \\
      & \texttt{s1c-s6-raw-s-vv-20250405t060936-20250405t060953-001754-00317f.zarr} \\
      & \texttt{s1c-s6-raw-s-vv-20250422t061414-20250422t061446-002002-0040d4.zarr} \\
      & \texttt{s1c-s6-raw-s-vv-20250422t061439-20250422t061511-002002-0040d4.zarr} \\
      & \texttt{s1c-s6-raw-s-vv-20250422t061504-20250422t061520-002002-0040d4.zarr} \\
      & \texttt{s1c-s6-raw-s-vv-20250429t060847-20250429t060920-002104-0046e6.zarr} \\
    \midrule
    \multirow{8}{*}{Validation}
      & \texttt{s1c-s1-raw-s-vv-20250328t052810-20250328t052835-001637-002a0d.zarr} \\
      & \texttt{s1c-s2-raw-s-vv-20250331t205042-20250331t205107-001690-002d5c.zarr} \\
      & \texttt{s1c-s3-raw-s-vv-20250414t083258-20250414t083323-001887-0039f0.zarr} \\
      & \texttt{s1c-s4-raw-s-vv-20250330t051150-20250330t051214-001666-002bd3.zarr} \\
      & \texttt{s1c-s4-raw-s-vv-20250407t204234-20250407t204258-001792-0033e2.zarr} \\
      & \texttt{s1c-s5-raw-s-vv-20250406t050339-20250406t050403-001768-003264.zarr} \\
      & \texttt{s1c-s5-raw-s-vv-20250420t160613-20250420t160637-001979-003f6e.zarr} \\
      & \texttt{s1c-s6-raw-s-vv-20250401t161424-20250401t161448-001702-002e19.zarr} \\
    \midrule
    \multirow{8}{*}{Test}
      & \texttt{s1c-s1-raw-s-vv-20250424t170801-20250424t170839-002038-0042df.zarr} \\
      & \texttt{s1c-s2-raw-s-vv-20250405t060506-20250405t060539-001754-00317b.zarr} \\
      & \texttt{s1c-s4-raw-s-vv-20250328t145502-20250328t145535-001643-002a6d.zarr} \\
      & \texttt{s1c-s4-raw-s-vv-20250328t145527-20250328t145555-001643-002a6d.zarr} \\
      & \texttt{s1c-s4-raw-s-vv-20250422t172412-20250422t172445-002009-00413d.zarr} \\
      & \texttt{s1c-s4-raw-s-vv-20250426t023204-20250426t023237-002058-004414.zarr} \\
      & \texttt{s1c-s5-raw-s-vv-20250421t022419-20250421t022447-001985-003fc0.zarr} \\
      & \texttt{s1c-s5-raw-s-vv-20250421t145445-20250421t145518-001993-004045.zarr} \\
    \bottomrule
  \end{tabular}
\end{table*}

\section{Extended Discussion}
\label{app:discussion}

\subsection{Interpretation}
The central result of this paper is that, after standard front-end preprocessing, a substantial portion of the azimuth-focusing operator can be represented as a compact causal dynamical system. This explains why streamAZ can reduce latency and memory so aggressively without collapsing reconstruction quality: the model does not try to learn all of SAR imaging end to end, but instead targets the specific long-range phase-history transformation that remains after range-domain preparation.

The experiments also clarify which design choices matter most. Performance improvements did not come primarily from making the teacher wider or deeper; rather, they came from regularization and from balancing losses that preserve amplitude structure, sharpness, and spectral content. This is consistent with the intuition that online focusing is less a brute-force function-approximation problem than a structurally constrained operator-learning problem.

\subsection{Complexity and FLOP estimates}

To highlight the advantages of our Online Processor, we compare our proposed method against two implementations of the RDA algorithm, one being the commonly used "batched" RDA where all operations are performed in batches, and the other is a version of the RDA which relies on buffering a window of SAR history that can perform focusing of a single range line at a time that we call the linewise RDA. We calculate that to be able to make an accurate focusing of a single line of SAR data, the linewise RDA must maintain a rolling buffer of 972 range lines, i.e. 486 range lines ahead of the line of interest and 486 lines behind the line of interest.

For our computational complexity estimation we assume that the values for RC, RCMC and AC filters are calculated prior to the beginning of the SAR data collection and thus ignore these contributions to computational cost. The reason for this is in the application of live SAR focusing we care only about computational cost which occurs between the points of collecting raw data and outputting focused data that contributes to the delay in obtaining focused data.

\begin{table*}[h]
  \centering
  \caption{Complexity, latency, memory, and compute comparison. Lower latency, memory, and GFLOPs per row are better.}
  \label{tab:complexity-comparison}
  \scriptsize
  \setlength{\tabcolsep}{3pt}
  \renewcommand{\arraystretch}{1.1}
  \begin{tabular}{@{}p{0.15\linewidth}p{0.20\linewidth}ccccc@{}}
    \toprule
    \textbf{Method} & \textbf{Complexity} & \textbf{Single-line} & \textbf{Full scan} & \textbf{Memory} & \textbf{GFLOPs} & \textbf{GFLOPs} \\
    &  & \textbf{latency (ms)}$^{*}$ & \textbf{latency (ms)}$^{*}$ & \textbf{footprint} & \textbf{per scan} & \textbf{per row} \\
    \midrule
    Traditional RDA & $\mathcal{O}\!\left(N_a N_r \log(N_a N_r)\right)$ & N/A$^{\dagger}$ & \textbf{27,444} & 16 GB & \textbf{122} & N/A$^{\dagger}$ \\
    Linewise RDA & $\mathcal{O}\!\left(N_a N_r N_b \log(N_b N_r)\right)$ & 857 (\textit{$+243$})$^{\ddagger}$ & 17,140,000 & 780 MB & 45,400 & 2.27 \\
    \rowcolor{gray!12}
    \textit{This work} & {\boldmath$\mathcal{O}(N_a N_r \log(N_r))$} & \textbf{15.1} & 302,000 & \textbf{6 MB} & 324 & \textbf{0.016} \\
    \bottomrule
  \end{tabular}
  \vspace{2pt}

  {\footnotesize $^{*}$Latency measured on a single core of an AMD EPYC 7343 CPU.}\\
  {\footnotesize $^{\dagger}$Cannot be used to process individual range lines in an online manner.}\\
  {\footnotesize $^{\ddagger}$Processing starts only after the history buffer is filled, adding delay beyond pure compute time.}
\end{table*}

Due to the fact our model uses fixed memory footprint SSM layers as its backbone and does not require buffering data in the azimuth direction, we obtain a far superior memory footprint to both baselines. Our online formulation enables drastically reduced data-in-to-data-out focusing latency; however, accumulated over an entire scan, the computational complexity of our technique is worse and that is also reflected in full scan processing latency. We show detailed computational complexity and FLOPs derivations in Appendices ~\ref{app:complexity} and ~\ref{app:flops}.

\subsection{Limitations and failure cases}
\label{app:limits-and-failure}

Our study is intentionally narrow in operating regime. The current formulation assumes a linear synthetic aperture: stripmap acquisition, broadside geometry, modest angular extent, and front-end preprocessing that already makes the azimuth task well conditioned. The deployed model is therefore not yet a drop-in replacement for wide-squint, spotlight, or heavily motion-corrupted collections.

Failure cases are most likely when those assumptions are violated. Large residual motion errors, stronger-than-expected range migration, unseen PRF or bandwidth regimes, and scene statistics far from Maya4 can all cause defocus, amplitude bias, or temporal lag in the emitted rows. In addition, because streamAZ is causal, it cannot exploit arbitrarily large future context in the way a fully offline batch imager can, so some loss of fidelity relative to the best offline methods should be expected in challenging scenes.

\subsection{Operational considerations}
\label{app:operational-considerations}
Operational \gls{sar} focusing in real systems typically requires motion compensation (\gls{imu}/\gls{gnss} + residual corrections), sidelobe control (windowing, multi-looking), and radiometric/geometric calibration.
Residual phase errors (platform motion, atmosphere/ionosphere, uncompensated squint) are commonly addressed with robust data-driven autofocus, with \gls{pga} remaining a strong baseline in many pipelines~\cite{wahl1994pga,cumming2005digital}.
These considerations highlight why deployment constraints matter for the proposed method. In practice, the value of online focusing is not only reduced FLOPs, but the ability to move interpretable image evidence earlier in the decision loop. A processor that emits focused rows incrementally allows downstream modules such as CFAR detection, target cueing, or adaptive mode selection to begin operating before a full aperture has been collected.

At the same time, streamAZ should be viewed as one component in a larger operational chain rather than as a complete replacement for classical estimation machinery. Motion compensation, autofocus, calibration, and uncertainty monitoring remain necessary, and in high-assurance settings the online product may be best used for rapid triage while slower high-fidelity imaging produces archival outputs.

\subsection{Sensitivity and robustness}
\label{app:sensitivity-robustness}
The teacher sweeps suggest that the method is more sensitive to objective design than to raw model size. Moderate changes to dropout and to the amplitude-correlation/edge balance produced larger gains than simply increasing capacity, and the final frontier showed a meaningful trade-off between balanced structural fidelity (ridge family) and RMSE specialization (selectivity family). This is encouraging for deployment because it suggests that good operating points can be found without dramatically increasing compute.

Robustness, however, is not fully characterized by the current experiments. The present study uses a single benchmark family and configuration sweeps rather than repeated-seed uncertainty estimates or explicit domain-shift stress tests. A stronger robustness claim would require evaluation across broader sensing conditions, perturbations of acquisition parameters, and downstream closed-loop tasks in which latency, not only image similarity, is measured directly.